\documentclass[12pt, a4paper]{article}
\usepackage{amsmath,amssymb,bm,epsfig,afterpage}
\usepackage{cite}
\usepackage{here}
\usepackage{color}
\usepackage{colortbl}
\usepackage{xcolor}
\usepackage{tabularx}
\usepackage{subcaption}
\usepackage{bbm}
\usepackage{graphicx}
\usepackage{listings}
\usepackage{longtable}
\usepackage[utf8]{inputenc}
\usepackage{cancel}
\usepackage{slashed}

\allowdisplaybreaks[3]

\newcommand{\lae}{{\lambda_e}}
\newcommand{\lpe}{{\lambda^\prime_e}}
\newcommand{\lan}{{\lambda_n}}
\newcommand{\lpn}{{\lambda^\prime_n}}

\newcommand{\Zp}{{Z^\prime}}
\newcommand{\gp}{{g^\prime}}
\newcommand{\Qp}{Q^\prime}
\newcommand{\ve}{\mathbf{e}}

\newcommand{\vn}{\mathbf{n}}

\newcommand{\GeV}{\mathrm{GeV}}

\newcommand{\la}{\lambda}
\newcommand{\eps}{\epsilon}

\newcommand{\tm}{\widetilde{m}}
\newcommand{\Mcal}{\mathcal{M}}
\newcommand{\Lcal}{\mathcal{L}}

\newcommand{\Ncal}{\mathcal{N}}

\newcommand{\Ecal}{\mathcal{E}}

\newcommand{\ev}{\mathbf{e}}

\newcommand{\nv}{\mathbf{n}}

\newcommand{\SM}{\mathrm{SM}}

\newcommand{\ol}[1]{\overline{#1}}

\newcommand{\vev}[1]{\langle{#1}\rangle}
\newcommand{\abs}[1]{\left|{#1}\right|}
\newcommand{\order}[1]{\mathcal{O}\left({#1}\right)}
\newcommand{\br}[2]{\mathrm{BR} \left({#1}\to{#2}\right)}

\newcommand{\id}[1]{\mathbf{1}_{{#1}} }

{\newcommand{\rep}[1]{\mathbf{#1}}

\newcommand{\Gam}[2]{\Gamma\left({#1}\to{#2}\right)}
\newcommand{\FB}{{\mathrm{FB}}}

\newcommand{\inv}{\mathrm{inv}}

\newcommand{\tc}{\tilde{c}}
\newcommand{\ts}{\tilde{s}}
\newcommand{\tM}{\widetilde{M}}

\newcommand{\PDG}{\mathrm{PDG}}
\newcommand{\CDF}{\mathrm{CDF}}

\usepackage[colorlinks=true, linkcolor=black, citecolor=black,
urlcolor=black]{hyperref}

\setcounter{footnote}{0}

\begin{document}

\begin{titlepage}

\begin{flushright}
 {\tt
CTPU-PTC-22-10
}
\end{flushright}

\vspace{1.2cm}
\begin{center}
{\Large
{\bf
$W$ mass in a model with vector-like leptons and $U(1)^\prime$
}
}
\vskip 2cm

Junichiro Kawamura$^{a}$~\footnote{jkawa@ibs.re.kr}
and
Stuart Raby$^b$~\footnote{raby.1@osu.edu}

\vskip 0.5cm

{\it $^a$
Center for Theoretical Physics of the Universe, Institute for Basic Science (IBS),
Daejeon 34051, Korea
}\\[3pt]

{\it $^b$
Department of Physics, Ohio State University, Columbus, Ohio, 43210, USA}\\[3pt]

\vskip 1.5cm

\begin{abstract}
We study the effects of vector-like leptons on the $W$ boson mass
in a model with a vector-like $U(1)^\prime$ gauge symmetry.
This model provides simultaneous explanations
for the recent anomalies in the muon anomalous magnetic moment
and the semi-leptonic decays of $B$ mesons.
We found that the recent result of the $W$ boson mass precise measurement
at CDF can be explained
if the charged (neutral) vector-like lepton is lighter than 250 (80) GeV.
The light vector-like leptons may not be excluded by collider experiments
if these decay to a physical mode of the $U(1)^\prime$ breaking scalar field.
\end{abstract}
\end{center}
\end{titlepage}

%\tableofcontents
\clearpage

\section{Introduction}

Recently, the CDF collaboration reported a new result
of the precise $W$ boson mass measurement~\cite{CDF:2022hxs},
\begin{align}
 m_W^\CDF = 80.4335~(94)~\GeV.
\end{align}
This value is larger than the combination of the previous measurements
$m_W^\PDG = 80.379~(12)~\GeV$
and the Standard Model (SM) prediction
$m_W^\SM = 80.361~(6)~\GeV$~\cite{ParticleDataGroup:2020ssz}.
Since the announcement of the result,
the explanations for the new $W$ boson mass and its relations to other physics
have been studied extensively~\cite{Strumia:2022qkt,deBlas:2022hdk,Yang:2022gvz,Yuan:2022cpw,Athron:2022qpo,Lu:2022bgw,Fan:2022dck,Babu:2022pdn,Heckman:2022the,Gu:2022htv,Athron:2022isz,DiLuzio:2022xns,Asadi:2022xiy,Bahl:2022xzi,Paul:2022dds,Bagnaschi:2022whn,Cheng:2022jyi,Lee:2022nqz,Liu:2022jdq,Fan:2022yly,Sakurai:2022hwh,Balkin:2022glu,Biekotter:2022abc,Endo:2022kiw,Crivellin:2022fdf,Heo:2022dey,Han:2022juu,Ahn:2022xeq,Song:2022xts,Blennow:2022yfm,Cacciapaglia:2022xih,Tang:2022pxh,Zhu:2022tpr,Zheng:2022irz,Krasnikov:2022xsi,Arias-Aragon:2022ats,Du:2022brr,Du:2022pbp,Kawamura:2022uft,
Kanemura:2022ahw,Nagao:2022oin,Zhang:2022nnh,Carpenter:2022oyg,Popov:2022ldh,Chowdhury:2022moc,Borah:2022obi,Zeng:2022lkk,Du:2022fqv,Ghorbani:2022vtv,Bhaskar:2022vgk,Baek:2022agi,Cao:2022mif,Borah:2022zim,Batra:2022org,Lee:2022gyf,Almeida:2022lcs,Cheng:2022aau,Heeck:2022fvl,Abouabid:2022lpg,Batra:2022pej,Benbrik:2022dja,Cai:2022cti,Zhou:2022cql,Gupta:2022lrt,Wang:2022dte,Barman:2022qix,Kim:2022hvh,Kim:2022xuo,Dcruz:2022dao,Isaacson:2022rts,
Chowdhury:2022dps,Kim:2022zhj,Gao:2022wxk,Lazarides:2022spe, Chen:2022ocr 
}.

In this work, we point out that the shift of the $W$ boson mass
can be explained in a model with vector-like (VL) leptons and an extra $U(1)^\prime$
gauge symmetry which was proposed in Refs.~\cite{Kawamura:2019rth,Kawamura:2019hxp}.
The model provides a simultaneous explanation
for the anomalies in the muon anomalous magnetic moment, $g-2$,
and the semi-leptonic decays of the $B$ meson~\footnote{
Models with VL fermions and a $U(1)^\prime$ symmetry for the anomalies are proposed in Refs.~\cite{Allanach:2015gkd,Altmannshofer:2016oaq,Megias:2017dzd,Raby:2017igl,King:2017anf,Darme:2018hqg,Crivellin:2015mga,Crivellin:2018qmi,Crivellin:2020oup,Crivellin:2021rbq,Alguero:2022est}. Explanations for $\Delta a_\mu$ by VL leptons are proposed in e.g. Refs.~\cite{Czarnecki:2001pv,Kannike:2011ng,Dermisek:2013gta,Poh:2017tfo,Kawamura:2020qxo,Bai:2021bau}. 
}.
The recent FNAL measurement~\cite{Abi:2021gix}
confirmed the long-standing discrepancy of the muon $g-2$
between the experimental value~\cite{Bennett:2004pv} and the SM prediction~\cite{Aoyama:2020ynm,Aoyama:2012wk,Aoyama:2019ryr,Czarnecki:2002nt,Gnendiger:2013pva,Davier:2017zfy,Keshavarzi:2018mgv,Colangelo:2018mtw,Hoferichter:2019gzf,Davier:2019can,Keshavarzi:2019abf,Kurz:2014wya,Melnikov:2003xd,Masjuan:2017tvw,Colangelo:2017fiz,Hoferichter:2018kwz,Gerardin:2019vio,Bijnens:2019ghy,Colangelo:2019uex,Blum:2019ugy},
and the current discrepancy is $\Delta a_\mu = 2.51~(59)\times 10^{-9}$.
Yet another discrepancy from the SM is found in the measurements of rare semi-leptonic
$B$ meson decays~\cite{Aaij:2014ora,Aaij:2017vbb,
Aaij:2019wad,Abdesselam:2019wac,Aaij:2013aln,
Lees:2013nxa,Aaij:2014pli,Aaij:2015esa,
Aaij:2013qta,Khachatryan:2015isa,Aaij:2015oid,Abdesselam:2016llu,Wehle:2016yoi,ATLAS:2017dlm,CMS:2017ivg,Aaij:2020ruw,Aaij:2021vac}, $b\to s\ell\ell$.
In this paper, we focus on the VL leptons and $\Zp$ boson in this model
to show that $m_W$ and $\Delta a_\mu$ can be explained simultaneously.
As we have shown in Refs.~\cite{Kawamura:2019rth,Kawamura:2019hxp},
this model can easily accommodate the $b\to s\ell\ell$ anomaly
when the $\Zp$ boson is sufficiently light and strongly couples to muons,
so that $\Delta a_\mu$ is explained.
Therefore, our model will provide a unified explanation
for the recent three anomalies, $m_W$, $\Delta a_\mu$ and $b\to s\ell\ell$.
It will turn out that the VL leptons should be 
lighter than those to explain $\Delta a_\mu$. 
Hence we will study the observables 
which would be changed from the SM values due to the light VL leptons.

The rest of this paper is organized as follows.
The model is briefly introduced in Sec.~\ref{sec-model},
and then the observables in our analysis are discussed in Sec.~\ref{sec-pheno}.
The result of a numerical analysis is shown
and the LHC signals are discussed in Sec.~\ref{sec-rslt}.
Section~\ref{sec-concl} is devoted to summary.
Diagonalization of the mass matrices is discussed in Appendix~\ref{sec-nonzero},  
and the formula for the three-body decays are shown in Appendix~\ref{sec-three}.

\section{Model}
\label{sec-model}

\subsection{Mass matrices}

We briefly introduce our model,
with particular interests in VL leptons,
see Refs.~\cite{Kawamura:2019rth,Kawamura:2019hxp} for more details.
The matter content relevant to the discussion in this work
is shown in Table~\ref{tab-MTCT}.
We assume that the VL leptons mix with only the SM leptons in the second generation
to prevent flavor violations.
The mass terms and the Yukawa couplings are given by
\begin{align}
 \Lcal \supset&\ - m_L \ol{L}_R L_L -  m_E \ol{E}_R E_L - m_N \ol{N}_R N_L
 - \frac{1}{2} M_{R} \ol{\nu}_R^c \nu_R   \\ \notag
         &\
 + y_2 \ol{\mu}_R \ell_L H  + y_n \ol{\nu}_R \ell_L \tilde{H}
 + \la_L \Phi   \ol{L}_R \ell_L- \la_E \Phi^* \ol{\mu}_R E_L - \la_N \Phi^* \ol{\nu}_R N_L
 \\  \notag
 &\
     + \lae \ol{E}_R L_L H  - \lpe \ol{L}_R \tilde{H} E_L
       + \lan \ol{N}_R L_L \tilde{H}  + \lpn \ol{L}_R {H} N_L
                + h.c. ,
\end{align}
where $\tilde{H} := i\sigma_2 H^* = (H_-^*, -H_0^* )$.
The $SU(2)_L$ indices are contracted by $i\sigma_2$.
We introduce the Majorana mass term of $\nu_R$ for the type-I see-saw mechanism.
After the scalar fields develop their vacuum expectation values (VEVs),
$v_H := \vev{H_0}$ and $v_\Phi := \vev{\Phi}$,
the Dirac mass matrices for the leptons are given by
\begin{align}
\label{eq-Me}
 \ol{\ve}_R \Mcal_e \ve_L  :=&\
\begin{pmatrix}
 \ol{\mu}_R & \ol{E}_R & \ol{E}_R^\prime
\end{pmatrix}
\begin{pmatrix}
 y_2 v_H & 0        & \la_E v_\Phi \\
 0         & \lae v_H & m_E \\
\la_L v_\Phi & m_L & \lpe v_H
\end{pmatrix}
\begin{pmatrix}
 \mu_L \\ E^\prime_L \\ E_L
\end{pmatrix},  \\
\ol{\vn}_R \Mcal_n \vn_L := &\
\begin{pmatrix}
 \ol{\nu}_R & \ol{N}_R & \ol{N}_R^\prime
\end{pmatrix}
\begin{pmatrix}
 y_n v_H         & 0        & \la_N v_\Phi \\
 0         & \lan v_H & m_N \\
\la_L v_\Phi & m_L & \lpn v_H
\end{pmatrix}
\begin{pmatrix}
 {\nu}_L \\ {N}^\prime_L \\ N_L
\end{pmatrix}.
\label{eq-Mn}
\end{align}

\begin{table}[t]
\centering
 \caption{\label{tab-MTCT}
Matter contents.
Electric charge of fermion $f$ is $Q_f = T_f^3+Y_f/2$.
The $SU(2)_L$ doublets have components;
$\ell_L = (\nu_L, \mu_L)$, $H=(H_0,H_-)$, $L_L = (N_L^\prime, E_L^\prime)$
and $\ol{L}_R = (-\ol{E}_R^\prime, \ol{N}_R^\prime)$.
}
\begin{tabular}[t]{c|cccc|cccccc|cc} \hline
              & $\ell_L$ & $\ol{\mu}_R$ & $\ol{\nu}_R$ & $H$ & $L_L$ & $\ol{E}_R$ & $\ol{L}_R$ & $E_L$
              & $N_L$ & $\ol{N}_R$
              & $Z^\prime$ & $\Phi$ \\ \hline \hline
$SU(2)_L$     & $\rep{2}$& $\rep{1}$ & $\rep{1}$ & $\rep{2}$& $\rep{2}$&$\rep{1}$ &$\rep{2}$ &$\rep{1}$ & $\rep{1}$& $\rep{1}$ & $\rep{1}$ & $\rep{1}$ \\
$U(1)_Y$      & $-1$     & $2$    & $0$  & $-1$     & $-1$      & $2$   & $1$ & $-2$ & $0$ & $0$  &$0$ & $0$\\
$U(1)^\prime$ & $0$      & $0$   & $0$    & $0$      & $-1$      & $1$      &$1$ & $-1$
& $-1$ & $1$ & $0$ & $-1$ \\
\hline
\end{tabular}
\end{table}

The mass basis is defined as
\begin{align}
 \hat{\ve}_L := U_{e_L}^\dag \ve_L,
\quad
 \hat{\ve}_R := U_{e_R}^\dag \ve_R,
\quad
 \hat{\vn}_L := U_{n_L}^\dag \vn_L,
\quad
 \hat{\vn}_R := U_{n_R}^\dag \vn_R,
\end{align}
where unitary matrices diagonalize the mass matrices as
\begin{align}
 U_{e_R}^\dag \Mcal_e U_{e_L} =
\begin{pmatrix}
 m_\mu & 0 & 0 \\ 0 & m_{E_1} & 0 \\ 0 & 0 & m_{E_2}
\end{pmatrix}, 
\quad
U_{n_R}^\dag \Mcal_n U_{n_L} =
\begin{pmatrix}
 \cdot & \cdot & \cdot \\ 0 & m_{N_1} & 0 \\ 0 & 0 & m_{N_2}
\end{pmatrix}.
\end{align}
The masses are increasingly ordered.
Here,
$\cdot$'s in the first row of the neutrino mass matrix
will be irrelevant  after $\nu_R$ is integrated out.
Neglecting $\order{v_H}$ entries,
the VL lepton masses are given by
\begin{align}
 m_{E_1} \sim M_E := \sqrt{m_E^2 + \la_E^2 v_\Phi^2},
\quad
m_{N_1} \sim m_N,
\quad
 m_{E_2} \sim m_{N_2} \sim M_L := \sqrt{m_L^2 + \la_L^2 v_\Phi^2},
\end{align}
when $M_E, m_N < M_L$ and hence $L \sim (E_2, N_2)$.

We define the Dirac fermions as
\begin{align}
\hat{\ev} := \left(\mu, E_1, E_2\right),  
\quad
\hat{\nv} := \left(\nu, N_1, N_2 \right),
\end{align}
 where
\begin{align}
\left[\ev\right]_i := \left( \left[\hat{\ev}_L\right]_i,
                          \left[\hat{\ev}_R \right]_i \right),
\quad
\left[\nv\right]_i := \left( \left[\hat{\nv}_L\right]_i,
                          \left[\hat{\nv}_R \right]_i \right),
\end{align}
with $i=1,2,3$.
We expand the neutral scalar fields as
\begin{align}
 H_0 = v_H + \frac{1}{\sqrt{2}} \left(h + i a_h\right),
\quad
 \Phi = v_\Phi + \frac{1}{\sqrt{2}} \left(\chi + i a_\chi\right),
\end{align}
where $h$ and $\chi$ are the physical real scalar fields,
while the pseudo-scalar components $a_h$ and $a_\chi$ are absorbed
by the $Z$ and $\Zp$ bosons, respectively.

\subsection{Interactions}
The $Z$ and $W$ boson couplings are given by
\begin{align}
 \Lcal_{Z,W} =&\ Z_\mu \sum_{f=\ve,\vn} \ol{f} \gamma^\mu
 \left(g^Z_{f_L} P_L + g^Z_{f_R} P_R \right)   f
 + \Bigl[  W_\mu \ol{\vn}
 \gamma^\mu  \left({g}^W_{L} P_L + {g}^W_{R} P_R \right) \ve + h.c. \Bigr],
\end{align}
where
\begin{align}
 g^Z_{\ve_A} =  \frac{g}{2 c_W} \left(
    - \Ecal^A 
   + 2 s_W^2 \id{3}\right),
\quad
 g^Z_{\vn_A} =  \frac{g}{2 c_W} \Ncal^A, 
\quad
 g^W_{A} =  \frac{g}{\sqrt{2}} h^A, 
\end{align}
for $A=L,R$.
Here,
\begin{align}
\Ecal^A := U_{e_A}^\dag Q_A U_{e_A},
\quad
\Ncal^A :=  U_{n_A}^\dag Q_A U_{n_A},
\quad
h^A := U_{n_A}^\dag Q_A U_{e_A},
\end{align}
where  $Q_L=\mathrm{diag}(1,1,0)$ and $Q_R = \id{3}-Q_L$.
$P_{L}$ ($P_R$) are the chiral projections onto the left- (right-) handed fermions.

The gauge interactions with the $\Zp$ boson in the mass basis are defined as
\begin{align}
 \Lcal_{\Zp} =&\ Z^\prime_\mu \sum_{f=\ve, \vn}
                  \ol{f} \gamma^\mu \left({g}^\Zp_{f_L} P_L + {g}^\Zp_{f_R}P_R  \right)
                  f,
\end{align}
where the coupling matrices are given by
\begin{align}
g^\Zp_{\ev_A} =&\ \gp U_{e_A}^\dag \Qp_e
  U_{e_A}, \quad
g^\Zp_{\nv_A} = \gp U_{n_A}^\dag \Qp_n U_{n_A},
\end{align}
with $\Qp_e = \Qp_n = \mathrm{diag}(0,-1,-1)$.
$\gp$ is the gauge coupling constant for $U(1)^\prime$.

The Yukawa interactions are given by
\begin{align}
-\Lcal_Y = \frac{1}{\sqrt{2}}
      \sum_{S=h, \chi}  \sum_{f=\ve,\vn} S \ol{f}\; Y^S_f P_L f + h.c. ,
\end{align}
where
\begin{align}
 Y^h_\ve =&\ U_{e_R}^\dag
\begin{pmatrix}
y_2 & 0 & 0 \\ 0 & \lae & 0 \\ 0 & 0 & \lpe  		
\end{pmatrix}
U_{e_L},
& \quad
 Y^\chi_\ve =&\ U_{e_R}^\dag
\begin{pmatrix}
0 & 0 & \la_E \\ 0 & 0 & 0 \\ \la_L & 0 & 0
\end{pmatrix}
U_{e_L},
\\
Y^h_\vn =&\  U_{n_R}^\dag
\begin{pmatrix}
y_n & 0 & 0 \\ 0 & \lan & 0 \\ 0 & 0 & \lpn  		
\end{pmatrix}
U_{n_L}
,
&\quad
Y^\chi_\vn =&\ U_{n_R}^\dag
\begin{pmatrix}
 0 & 0 & \la_N \\ 0 & 0 & 0 \\ \la_L & 0& 0
\end{pmatrix}
U_{n_L}.
\end{align}

\subsection{Couplings at the leading order}

We show the structures of the coupling matrices at the leading order
in $\order{m_\mu/v_\Phi}$.
Hereafter, we assume $\la_e, \la_n \sim y_2$, 
so that the muon mass is explained without fine-tuning, 
and the $Z$ and $W$ boson couplings to the SM leptons
do not sizably deviate from the SM values.  
The formulas with the sub-leading terms 
are explicitly shown in Appendix~\ref{sec-nonzero}.
The $\Zp$ couplings  are given by
\begin{align}
g^\Zp_{\ve_L} =\gp  U_L^\dag Q_L^\prime U_L \sim   &\ -\gp
\begin{pmatrix}
 s_L^2 & - c_L s_L c_{e_L} & c_L s_L s_{e_L} \\
- c_L s_L c_{e_L} & s_{e_L}^2 + c_L^2 c_{e_L}^2 & s_L^2 s_{e_L} c_{e_L} \\
 c_L s_L s_{e_L} & s_L^2 s_{e_L} c_{e_L}& c_L^2 s_{e_L}^2 + c_{e_L}^2
\end{pmatrix},
\\
g^\Zp_{\ve_R} =
\gp U_R^\dag Q_R^\prime U_R \sim   &\ -\gp
\begin{pmatrix}
 s_E^2 & - c_E s_E s_{e_R} & - c_E s_E c_{e_R} \\
- c_E s_E s_{e_R} & c_{e_R}^2 + c_E^2 s_{e_R}^2 & - s_E^2 s_{e_R} c_{e_R}  \\
- c_E s_E c_{e_R} & - s_E^2 s_{e_R} c_{e_R} & c_E^2 c_{e_R}^2 + s_{e_R}^2
\end{pmatrix},
\end{align}
where
\begin{align}
 c_X := \frac{m_X}{M_X},
\quad
 s_X := \frac{\la_X v_\Phi}{M_X},
\quad
 X =L, E.
\end{align}
Here, $c_{e_A}$ and $s_{e_A}$ ($A=L,R$) are the angles to diagonalize
the VL mass matrix,   defined in Eq.~\eqref{eq-cseLR}.
These are approximately given by
\begin{align}
 c_{e_A} \simeq 1-\frac{\eta_{e_A}^2}{2},
\quad
 s_{e_A} \simeq \eta_{e_A},
\quad
\eta_{e_L} := \frac{\lpe v_H M_L}{M_L^2 - M_E^2},
\quad
\eta_{e_R} := \frac{\lpe v_H M_E}{M_L^2 - M_E^2},
\end{align}
for $\eta_{e_A} \ll 1$.
Those for the neutrino couplings $g^{\Zp}_{\vn_A}$ 
are given by replacing $s_E \to 0$, $c_E \to1$ and $e_A\to n_A$.
The isospin parts of the $Z$ and $W$ boson couplings are given by
\begin{align}
 \Ecal^A
\sim
\begin{pmatrix}
 \delta_{AL} & 0 & 0 \\
0 & c_{e_A}^2 & - c_{e_A}s_{e_A} \\
0 & -c_{e_A} s_{e_A} &  s_{e_A}^2 \\
\end{pmatrix},
\quad
 \Ncal^A
\sim
\begin{pmatrix}
 \delta_{AL} & 0 & 0 \\
0 & c_{n_A}^2 & - c_{n_A}s_{n_A} \\
0 & -c_{n_A} s_{n_A} &  s_{n_A}^2 \\
\end{pmatrix},
\end{align}
and
\begin{align}
  h^A
\sim
\begin{pmatrix}
 \delta_{AL} & 0 & 0 \\
0 & c_{e_A}c_{n_A} & - c_{n_A}s_{e_A} \\
0 & -c_{e_A} s_{n_A} &  s_{e_A}s_{n_A} \\
\end{pmatrix},
\end{align}
for $A=L,R$. Here, $\delta_{AL} = 1~(0)$ for $A=L~(R)$.
Thus, the SM lepton couplings to the $Z$ and $W$ bosons
are the same as the SM ones,
and the off-diagonal couplings of the SM and VL fermions are vanishing,
up to $\order{m_\mu/v_\Phi}$.
This is in contrast to the model studied in Ref.~\cite{Lee:2022nqz},
and hence the $W$-mass could be addressed in this model solely by the VL leptons.

The Yukawa couplings to the scalars $h$ and $\chi$ are respectively given by
\begin{align}
Y^h_{\ve}
\sim &\
  \begin{pmatrix}
y_2 c_L c_E + \la_e s_L s_E & \order{y_2}  &  \order{y_2} \\
\order{y_2} & s_{e_L} c_{e_R} \lpe & c_{e_L} c_{e_R} \lpe \\
\order{y_2} & - s_{e_L}s_{e_R} \lpe & - c_{e_L}s_{e_R} \lpe
 \end{pmatrix},  \\
 Y^\chi_\ve
\sim &\
 \begin{pmatrix}
0 &  c_{E} s_{e_L} \la_E &  c_E c_{e_L} \la_E \\
c_L c_{e_R} \la_L
& s_E s_{e_L} s_{e_R}\la_E + s_L c_{e_L}  c_{e_R}\la_L  &
  s_E c_{e_L} s_{e_R}\la_E - s_L s_{e_L}  c_{e_R}\la_L  \\
- c_L s_{e_R}\la_L &
  s_E s_{e_L} c_{e_R}\la_E - s_L c_{e_L}  s_{e_R}\la_L  &
  s_E c_{e_L} c_{e_R}\la_E + s_L s_{e_L}  s_{e_R}\la_L  &
 \end{pmatrix}.
\label{eq-Ychie}
\end{align}
Those for the neutrinos are given by replacing $y_2\to 0$, $e\to n$ and $E\to N$.
Note that $m_\mu \sim( y_2 c_L c_E + \la_e s_L s_E) v_H$, 
so the Yukawa coupling to the Higgs boson
is also not changed from the SM value for heavy VL leptons. 
The SM Higgs couplings to the SM and VL leptons are again suppressed
by the small muon Yukawa coupling.
The $\chi$ boson coupling to the SM muon vanishes
if we neglect the muon mass.

\section{Phenomenology}
\label{sec-pheno}

\subsection{Oblique parameters and $W$ boson mass}
\label{sec-mW}

In this model, the VL lepton contributions to the $T$ parameter,
one of the oblique parameters~\cite{Peskin:1990zt,Peskin:1991sw},
is given by~\cite{Lavoura:1992np},
\begin{align}
16\pi s_W^2 c_W^2  T =&\  
\sum_{a,\beta} \left\{
 \left( \abs{h^L_{a\beta}}^2 + \abs{h^R_{a\beta}}^2\right)\theta_+(y_a, y_\beta)
      + 2\mathrm{Re}\left(h^L_{a\beta}h^{R*}_{a\beta}\right) \theta_-(y_a,y_\beta)
  \right\}
\\  \notag
&\ - \sum_{a<b} \left\{
 \left( \abs{\Ncal^L_{ab}}^2 + \abs{\Ncal^R_{ab}}^2\right)\theta_+(y_a, y_b)
      + 2\mathrm{Re}\left(\Ncal^L_{ab} \Ncal^{R*}_{ab}\right) \theta_-(y_a,y_b)
  \right\}  \\ \notag
&\ 
-\sum_{\alpha<\beta} \left\{
 \left( \abs{\Ecal^L_{\alpha\beta}}^2 + \abs{\Ecal^R_{\alpha\beta}}^2\right)\theta_+(y_\alpha, y_\beta)
      + 2\mathrm{Re}\left(\Ecal^L_{\alpha\beta}\Ecal^{R*}_{\alpha\beta}\right)
 \theta_-(y_\alpha,y_\beta)
  \right\},
\end{align}
where $a,b=1,2,3$ ($\alpha,\beta=1,2,3$) run over the neutral (charged) leptons. 
These subscripts are the elements of the couplings matrices, 
e.g. $h^L_{a\beta} := [h^L]_{a\beta}$. 
The other oblique parameters, $2\pi S$ and $-2\pi U$, are obtained by replacing
the functions, $\theta_\pm \to \chi_\pm$,
except for the first line in the $S$ parameter
which should be replaced as $\theta_\pm \to \psi_\pm$.
The loop functions are defined in Ref.~\cite{Kawamura:2022uft}.
For an order of magnitude estimation, taking $s_{e_A}, s_{n_A} = 0$ and $m_{L^0}\simeq m_{L^-}$, we have
\begin{align}
16\pi s_W^2 c_W^2  T
\sim \frac{2(m_{L^-}^2-m_{L^0}^2)^2}{3m_Z^2M_L^2 }
 \sim&\ \frac{2v_H^4}{3m_Z^2 M_L^2}
 \left( \frac{\lpe^2}{1-M_E^2/M_L^2} - \frac{\lpn^2}{1-m_N^2/M_L^2}\right)^2,
\end{align}
where $m_{L^-}$ and $m_{L^0}$
are the masses of doublet-like charged and neutral VL leptons, respectively.
$\eta_{e_A}, \eta_{n_A} \ll 1$ is assumed in the second equality.
Hence,
\begin{align}
\label{eq-esT}
 T \sim 0.13 \times
 \left( \frac{\lpe^2}{1-M_E^2/M_L^2} - \frac{\lpn^2}{1-m_N^2/M_L^2}\right)^2
  \left(\frac{250~\GeV}{M_L}\right)^2,
\end{align}
where we used $s_W^2 = 0.23121$~\cite{ParticleDataGroup:2020ssz}.
The $W$ boson mass
is related to the oblique parameters as~\cite{Maksymyk:1993zm,Grimus:2008nb}~\footnote{
We neglect the difference in the definitions of the oblique parameters.
The length of slopes of self-energies are taken to be finite value~\cite{Lavoura:1992np},
but these effects are less than $5\%$ in the parameter space studied in this paper,
and thus not important.
}
\begin{align}
  \frac{\delta m_W^2}{m_W^2|_\SM} =&\ \frac{\alpha_e}{c_W^2-s_W^2}
    \left(  -\frac{1}{2} S + c_W^2 T + \frac{c_W^2-s_W^2}{4s_W^2} U
 + \frac{s_W^2}{\alpha_e} \eps_\mu 
  \right)
 \sim 0.001 \times \left(\frac{T}{0.1}\right),
\end{align}
where $T\gg S,U$ and $\eps_\mu \simeq 0$ are assumed in the second equality.
We include the shift from the $W$ boson coupling to the muon at tree-level, 
$h^L_{11} =: h^L_{\nu\mu}=: 1 - \eps_\mu$.  
Hereafter, we write the indices for the SM leptons by $\nu/\mu$ 
instead of $1$, so that these are not confused with the first generations.   
Note that $\eps_\mu$ is positive in our model as shown in Eq.~\eqref{eq-UL1}. 
Therefore, from Eq.~\eqref{eq-esT}, the shift may be explained
if $\lpe, \lpn \sim \order{1}$, $M_L \sim 250~\GeV$
and $M_E^2/M_L^2, m_N^2/M_L^2$ are $\order{1}$. 
%the VL masses for singlets, $M_E$ and $m_N$, are not so larger than $M_L$. 

\subsection{Muon $g-2$}

The muon anomalous magnetic moment $\Delta a_\mu$
is shifted by the 1-loop effects via the $\Zp$ and $\chi$ bosons.
This is approximately  given by~\cite{Jegerlehner:2009ry,Dermisek:2013gta}
\begin{align}
\label{eq-delamu}
 \Delta a_\mu \sim&\ 
   - \frac{m_\mu \lpe v_H}{16\pi^2 m_{\Zp}^2} \tilde{C}_{LR},
\\ 
\tilde{C}_{LR} :=&\ 
\frac{\la_L \la_E m_L m_E}{M_L M_E}
\left(
 \frac{G_Z(x_L)-G_Z(x_E)}{x_L - x_E}
+ \frac{m_\Zp^2}{2m_\chi^2} \frac{{y_L} G_{{S}}(y_L)-{y_R} G_{{S}}(y_R)}{y_L-y_R}
\right),
\end{align}
where $x_L := M_L^2/m_\Zp^2$, $x_E:= M_E^2/m_\Zp^2$,
$y_L:= M_L^2/m_\chi^2$ and $y_E := M_E^2/m_\chi^2$.
The exact formula and the loop functions are shown
in Refs.~\cite{Kawamura:2019rth,Kawamura:2019hxp}.
The value of $\Delta a_\mu$ is estimated as
\begin{align}
 \Delta a_\mu \sim 5\times 10^{-9}
\times \lpe \left(\frac{500~\GeV}{m_\Zp}\right)^2
       \left(\frac{\tilde{C}_{LR}}{-0.01}\right).
\end{align}
Hence, the $\Zp$ boson with mass of order $500~\GeV$ can explain $\Delta a_\mu$.

\subsection{EW observables}

The mixing of the second generation leptons and the VL states 
can change the SM predictions of the EW boson couplings. 
The Fermi constant $G_\mu$ determined by the muon decay is given by 
\begin{align}
G_\mu = G_F \abs{h^L_{\nu\mu}}, 
\quad 
\frac{G_F}{\sqrt{2}} = \frac{g^2}{8 m_W^2},   
\end{align}
at the tree-level, where $G_\mu = 1.1663787\times 10^{-5}~\GeV^2$~\cite{ParticleDataGroup:2020ssz}.  
The partial width of the $W$ boson is given by 
\begin{align}
\br{W}{\mu\nu} =&\ \frac{g^2 m_W}{48 \pi \Gamma_W}  \abs{ h^L_{\nu\mu}}^2,
\end{align}
where the muon and neutrino masses are neglected.   
In the SM, the gauge coupling constant and $\Gamma_W$ are given by 
$g(m_Z)=0.65184~(18)$~\cite{Antusch:2013jca} and   
$\Gamma_W = 2.0895~(8)~\GeV$~\cite{Denner:1991kt}, respectively. 
We use these SM values for numerical analysis,    
since the branching fraction, 
$\propto g^2m_W/\Gamma_W$, is approximately independent of $g$ and $m_W$.

The $Z$ boson partial decay widths are given by 
\begin{align}
 \Gam{Z}{\mu\mu} =&\ \frac{G_Fm_Z^3}{12\sqrt{2}\pi} 
     \left( 1 + \frac{3 \alpha_e}{4\pi}\right)
     \left(\abs{-\Ecal^L_{\mu\mu} + 2s_W^2}^2 + \abs{-\Ecal^R_{\mu\mu}+2s_W^2 }^2 \right), 
\\ 
 \Gam{Z}{\inv} =&\ \frac{G_Fm_Z^3}{12\sqrt{2}\pi} 
     \left(2 + \abs{\Ncal^L_{\nu\nu}}^2 \right), 
\end{align}
where the $Z$ boson couplings to the electron and tau neutrinos 
are set to the SM values. 
The leading QED effect is included for $Z\to\mu\mu$. 
The asymmetry parameter $A_\mu$ and $A_{FB}^{\mu}$ are given by 
\begin{align}
 A_\mu = 
\frac{(\Ecal^{L}_{\mu\mu})^2 - (\Ecal^{R}_{\mu\mu} )^2 
     - 4 s_W^2 \left(\Ecal^L_{\mu\mu} - \Ecal^R_{\mu\mu}\right)}
     { (\Ecal^{L}_{\mu\mu})^2 + (\Ecal^{R}_{\mu\mu})^2  
    - 4 s_W^2 \left(\Ecal^L_{\mu\mu} + \Ecal^R_{\mu\mu}\right) 
      + 8 s_W^4}, 
\quad 
 A_{FB}^\mu = \frac{3}{4} A_e A_\mu. 
\end{align}
Since the $Z$ boson couplings to the electrons are the SM-like, 
$A_e = A_e^{\mathrm{SM}} = 0.1468~(03)$~\cite{ParticleDataGroup:2020ssz}. 
We shall use the SM value for the weak angle, $s_W^2 = 0.23155$~\cite{ParticleDataGroup:2020ssz} to calculate $A_\mu$.

Similarly to the SM gauge bosons, $h\to \mu\mu$ can deviate from the SM value.
We define the ratio of the width, 
\begin{align}
R_{\mu\mu} := \frac{\Gamma(h\to\mu\mu)}{\Gamma(h\to\mu\mu)_\SM} 
            = \frac{v_H^2}{m_\mu^2} \abs{ \left[Y^h_\ve\right]_{\mu\mu}}^2.  
\end{align} 
The decay rate of $h\to \gamma\gamma$ will deviate from the SM value
due to the loop effects mediated by the charged VL leptons.
We define the ratio of the width of $h\to\gamma\gamma$ to that in the SM
as %
\begin{align}
\label{eq-haa}
R_{\gamma\gamma} := \frac{\Gamma(h\to\gamma\gamma)}{\Gamma(h\to\gamma\gamma)_\SM}
=
\abs{ 1 + \sum_{i=1,2}
 \left( \left[Y^h_\ve\right]_{E_iE_i}
   \frac{v_H}{m_{E_i}}\right) \frac{A^H_{1/2}(\tau_{E_i})}{A_\SM}
}^2,
\end{align}
where $\tau_{I} = m_H^2/(4m_{I}^2)$ for the lower case $I= E_1, E_2$,
and the SM contribution $A_\SM$ is given by~\cite{Djouadi:2005gi}
\begin{align}
A_\SM :=
A_1^H (\tau_W) + \sum_{f}  N^f_c Q_f^2 A^H_{1/2}(\tau_{f}).
\end{align}
Here, $f$ runs over all the SM fermions.
$Q_f$ and $N_c^f$ are the electric charge and the number of colors of the fermion $f$,
respectively.
Assuming that the production cross sections are the same as in the SM, 
the current experimental values are 
$R_{\mu\mu} = 1.19\pm 0.34$
and 
$R_{\gamma\gamma} = 1.10\pm 0.07$~\cite{ParticleDataGroup:2020ssz}.

For $h^L_{\nu\mu} \ne 1$, the SM gauge boson couplings 
are not lepton flavor universal. 
We shall consider the decays of $Z$, $W$ bosons and $\tau$,
\begin{align}
\label{eq-LFUZW}
 \frac{\Gam{Z}{\mu\mu}}{\Gam{Z}{ee}} 
=&\ \frac{\abs{-\Ecal^L_{\mu\mu} + 2s_W^2}^2 + \abs{-\Ecal^R_{\mu\mu}+2s_W^2}^2}
       {1-4s_W^2 + 8 s_W^4},   \quad 
\quad 
 \frac{\Gam{W}{\mu\nu}}{\Gam{W}{e\nu}}  
= 
\abs{h^L_{\nu\mu}}^2, \\
 \frac{\Gam{\tau}{\mu\nu\nu}}{\Gam{\tau}{e\nu\nu}}  
=&\ \abs{h^L_{\nu\mu}}^2 F_\ell\left(\frac{m_\mu^2}{m_\tau^2}\right),  
\end{align} 
where 
\begin{align}
 F_\ell(y) = 1-8y+8y^3 - y^4 - 12 y^2\log y.  
\end{align}
We also study the ratio of the $Z/W$ boson decays to tau and muon, 
$\Gam{Z}{\tau\tau}/\Gam{Z}{\mu\mu}$ 
and 
$\Gam{W}{\tau\nu}/\Gam{W}{\mu\nu}$
which are given by the inverse of Eq.~\eqref{eq-LFUZW} in our model.

\subsection{Muon trident process}
The $\Zp$ boson can induce the so-called neutrino trident process,
$\nu_\mu N \to \nu_\mu \mu^+ \mu^- N$~\cite{Altmannshofer:2014cfa,Altmannshofer:2014pba,Magill:2016hgc,Ge:2017poy,Ballett:2018uuc,Altmannshofer:2019zhy}.
The cross section for this process at the CCFR experiment is estimated as~\cite{Altmannshofer:2019zhy,Zhou:2019vxt}
\begin{align}
R_\mathrm{CCFR} :=  \frac{\sigma_\text{CCFR}}{\sigma^\text{SM}_\text{CCFR}}
\simeq \frac{(1+4s_W^2 + \Delta g^V_{\mu\mu\mu\mu})^2 + 1.13 (1-\Delta g^A_{\mu\mu\mu\mu})^2 }
          {(1+4s_W^2 )^2 + 1.13 },
\end{align}
where the $\Zp$ boson contributions $\Delta g_{\mu\mu\mu\mu}^{V,A}$ are given by
\begin{align}
 \Delta g^{V,A}_{\mu\mu\mu\mu} =&\
 \frac{\sqrt{2}}{G_F\cdot 2 m_{Z'}^2} \left[ {g}^{Z'}_{\vn_L}\right] _{\nu_\mu\nu_\mu}
        \left(\left[ g^{Z'}_{\ve_R}\right] _{\mu\mu}
           \pm\left[g^{Z'}_{\ve_L}\right] _{\mu\mu}  \right)
\sim \frac{s_L^2(s_E^2 \pm s_L^2)}{2\sqrt{2} G_F v_\Phi^2}.
\end{align}
The experimentally observed rate is $\sigma_\text{CCFR}/\sigma^\text{SM}_\text{CCFR} = 0.82\pm 0.28$ at 95\% C.L~\footnote{
For the calculation of $R_\mathrm{CCFR}$,
we used $s_W^2 = 0.23129$ as in Ref.~\cite{Altmannshofer:2019zhy}.
}.

\subsection{$b\to s\mu\mu$ anomaly}
\label{sec-bsll}

We emphasize that the recent anomaly in the $b\to s\ell\ell$ process
can easily be explained in this model.
The Wilson coefficients for the semi-leptonic operators are given by
\begin{align}
 C_{9,10} \sim&\
 -\frac{\pi}{2\sqrt{2} \alpha_e G_F } \frac{1}{V_{tb}V_{ts}^*}
  \left( \frac{\la_E^2}{M_E^2} \pm \frac{\la_L^2}{M_L^2}\right) \frac{g^{\Zp}_{sb}}{\gp},
\end{align}
where $V_{tb}$ and $V_{ts}$ are the CKM elements,
and $g^\Zp_{sb}$ is the $\Zp$ boson coupling constant to $sb$
in the left-handed interaction.
The Wilson coefficient $C_9$ is estimated as
\begin{align}
 C_9 \sim -0.8 \times\left(\frac{200~\GeV}{m_\mathrm{VLL}}\right)^2
                     \left(\frac{g^{\Zp}_{sb}/\gp}{10^{-4}}\right),
\quad
\frac{1}{m_\mathrm{VLL}^2} :=  \frac{\la_E^2}{M_E^2} + \frac{\la_L^2}{M_L^2},
\end{align}
whereas the current favored values is $C_9 \in [-1.0, -0.5]$ depending
on the value of $C_{10}$~\cite{Aebischer:2019mlg,Alguero:2019ptt,Alok:2019ufo,Ciuchini:2019usw,Datta:2019zca,Kowalska:2019ley,Arbey:2019duh}.
Thus, the $b\to s\ell\ell$ anomaly can be explained
even with the small $\Zp$ coupling to quarks,
as long as those to the VL leptons
are sizable to explain the shift in $m_W$ and $\Delta a_\mu$.
With such small couplings to the $\Zp$ boson,
the flavor violations such as $B_s$-$\ol{B}_s$ mixing
will not deviate from the SM prediction due to the small couplings,
as we have explicitly shown in Refs.~\cite{Kawamura:2019rth,Kawamura:2019hxp}.
Further, the production cross sections at the LHC will be so small
that the di-muon signal is much below the current limit~\cite{Aad:2019fac}.

\subsection{Cabibbo angle anomaly} 

As pointed out in Refs.~\cite{Kirk:2020wdk,Crivellin:2020ebi}, 
the shift of the $W$ boson coupling to the muon could explain 
the recent Cabibbo angle anomaly, 
which may be caused by the disagreement between values $V_{us}$ 
determined from beta and Kaon decays. 
Let us consider the observable~\cite{Kirk:2020wdk}
\begin{align}
 R(V_{us}) := \frac{V_{us}^{K\mu}}
                    {\sqrt{1-|V_{ud}^\beta|^2- \abs{V_{ub}}^2}} 
           =  \frac{V_{us}}{\sqrt{1-|V_{ud}/h^L_{\nu\mu}|^2- \abs{V_{ub}}^2}} 
           \sim 1 + \abs{\frac{V_{ud}}{V_{us}}}^2 \eps_\mu. % \left(h^L_{\nu\mu}-1\right), 
\end{align}
where $V_{us}^{K\mu}$ ($V_{ud}^\beta$) is the value of the CKM element 
determined from the Kaon ($\beta$) decay. 
The CKM elements without superscript are those in our model 
which are assumed to be unitary~\footnote{
We neglect non-unitarity of the $3\times 3$ CKM matrix 
which could be induced by mixing with VL quarks,  
since we have shown that it is unitary up to $\order{10^{-11}}$ 
for typical cases~\cite{Kawamura:2019rth,Kawamura:2019hxp}. 
}. 
Here, $V_{us}^{K\mu} = 0.2252~(5)$, $V_{ub} = 4\times 10^{-3}$~\cite{ParticleDataGroup:2020ssz} and $V_{ud} = 0.97373~(09)$~\cite{Kirk:2020wdk}.  
The current measured value is $R(V_{us}) = 0.9891~(27)$~\cite{Kirk:2020wdk}. 
Since $\eps_\mu > 0$ in this model, 
the tension can not be resolved by the mixing with the VL leptons. 
We shall not include $R(V_{us})$ in our $\chi^2$ analysis 
because it can not be explained by the mixing with VL leptons, 
and could be explained by mixing with VL quarks.

\section{Numerical results and LHC signals}
\label{sec-rslt}

\subsection{$\chi^2$ fitting}

\begin{table}[t]
\centering 
\caption{\label{tab-ex}
The list of observables and their values studied in our $\chi^2$ analysis. 
}
\begin{tabular}[t]{c|cc|c} \hline 
  Obs.          & Exp.  & Unc.    & Remark \\ \hline \hline 
 $\Delta a_\mu\times 10^9$ & $2.51$ & $0.59$ & 
Refs.~\cite{Abi:2021gix,Bennett:2004pv,Aoyama:2020ynm,Aoyama:2012wk,Aoyama:2019ryr,Czarnecki:2002nt,Gnendiger:2013pva,Davier:2017zfy,Keshavarzi:2018mgv,Colangelo:2018mtw,Hoferichter:2019gzf,Davier:2019can,Keshavarzi:2019abf,Kurz:2014wya,Melnikov:2003xd,Masjuan:2017tvw,Colangelo:2017fiz,Hoferichter:2018kwz,Gerardin:2019vio,Bijnens:2019ghy,Colangelo:2019uex,Blum:2019ugy} \\ 
 $m_W$~[GeV]          & 80.4335& 0.0094 & CDF value~\cite{CDF:2022hxs}    \\  \hline 
$\br{W}{\mu\nu}$&  0.1063 & 0.0015 &  Ref.~\cite{ParticleDataGroup:2020ssz} \\
 $\Gam{Z}{\mu\mu}$~[GeV]&  0.08395& 0.00018&  Ref.~\cite{ParticleDataGroup:2020ssz} \\
 $\Gam{Z}{\inv}$~[GeV]  & 0.4989 & 0.0025  & Ref.~\cite{ParticleDataGroup:2020ssz}  \\
 $A_\mu$                & 0.142  & 0.015   & Ref.~\cite{ParticleDataGroup:2020ssz}  \\   
 $A_\FB^\mu$            & 0.0169 & 0.0013  & Ref.~\cite{ParticleDataGroup:2020ssz}  \\ 
 $R_{\mu\mu}$           & 1.19 & 0.34 & Ref.~\cite{ParticleDataGroup:2020ssz}  \\ 
 $R_{\gamma\gamma}$     & 1.10 & 0.07 & Ref.~\cite{ParticleDataGroup:2020ssz}  \\ \hline 
$\Gam{Z}{\mu\mu}/\Gam{Z}{ee}$ & 1.0001 & 0.0024    &  Ref.~\cite{ParticleDataGroup:2020ssz} \\ 
$\Gam{Z}{\tau\tau}/\Gam{Z}{\mu\mu}$ & 1.0010 & 0.0026    &  Ref.~\cite{ParticleDataGroup:2020ssz} \\ 
$\Gam{W}{\mu\nu}/\Gam{W}{e\nu}$ & 0.996 & 0.008    &  Ref.~\cite{ParticleDataGroup:2020ssz} \\ 
$\Gam{W}{\tau\nu}/\Gam{W}{\mu\nu}$ & 1.070 & 0.026    &  Ref.~\cite{ParticleDataGroup:2020ssz} \\ 
$\Gam{\tau}{\mu\nu\nu}/\Gam{\tau}{e\nu\nu}$ & 0.9762 & 0.0028  &  Ref.~\cite{ParticleDataGroup:2020ssz} \\ \hline 
 $R_{\mathrm{CCFR}}$    & 0.82 & 0.14 & Ref.~\cite{Altmannshofer:2019zhy,Zhou:2019vxt}\\
\hline
\end{tabular}
\end{table}

\newcommand{\fwidth}{0.49\textwidth}
\begin{figure}[t]
\centering
\includegraphics[width=\fwidth]{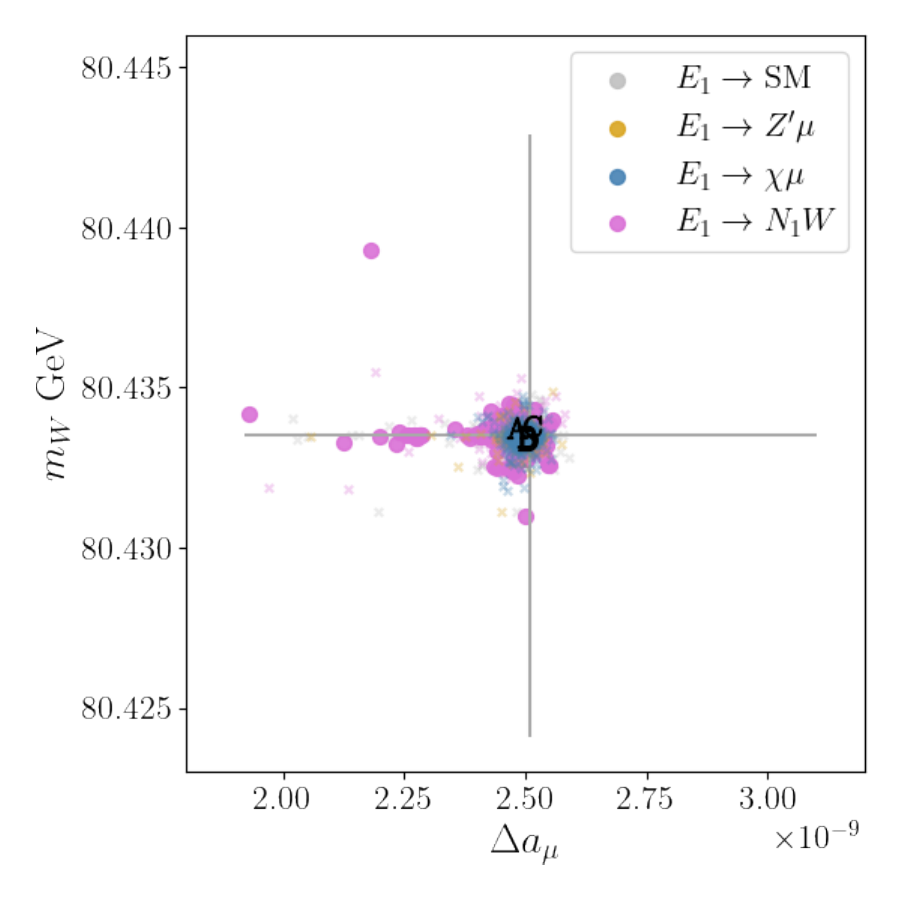}
\includegraphics[width=\fwidth]{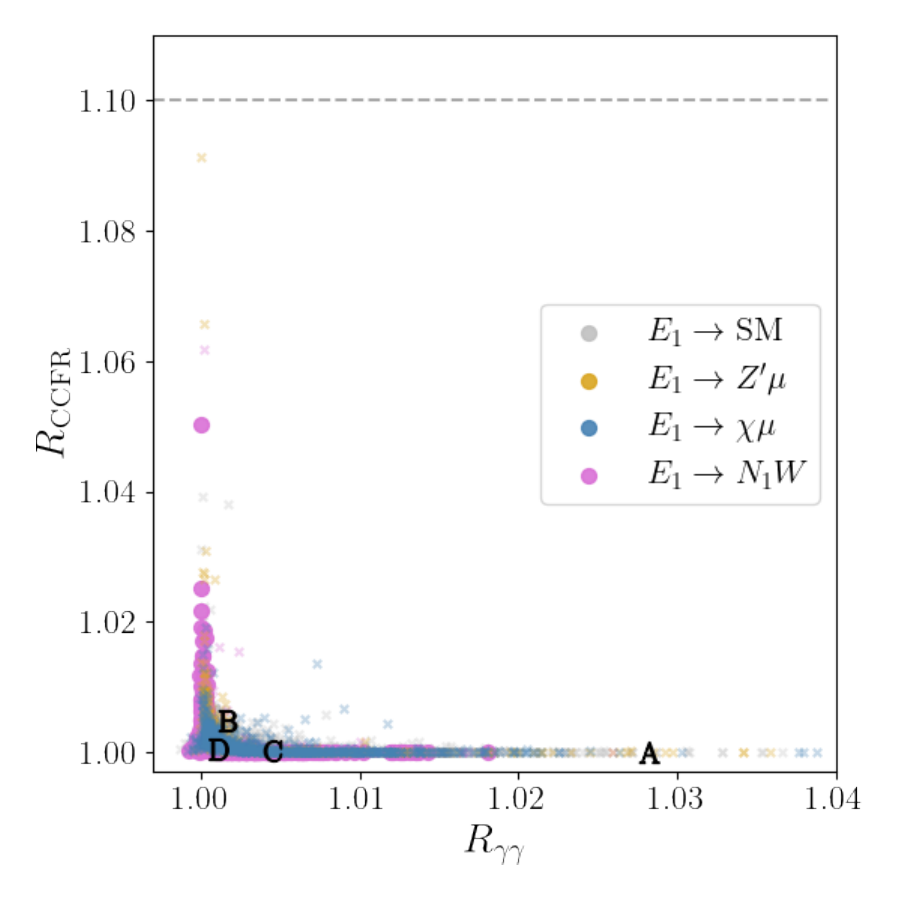}
\caption{\label{fig-obs}
Scattering plots on the observables in our analysis.
The cross and circle points are
$100 < m_{E_1} < 175~\GeV$ and $m_{E_1} > 175~\GeV$, respectively.
All the points explain the $m_W$ and $\Delta a_\mu$ within $1\sigma$. 
The error bars on the left panel are the $1\sigma$ uncertainties,
and the dash horizontal line on the right panel is 95\% C.L. limit on $R_\mathrm{CCFR}$.
The central value of $R_{\gamma\gamma} = 1.10$ is outside of the figure.  
The benchmark points, (A), (B), (C), and (D), lie on top of each other in the left panel.  
}
\end{figure}
\begin{figure}[tbh]
\centering
\includegraphics[width=\fwidth]{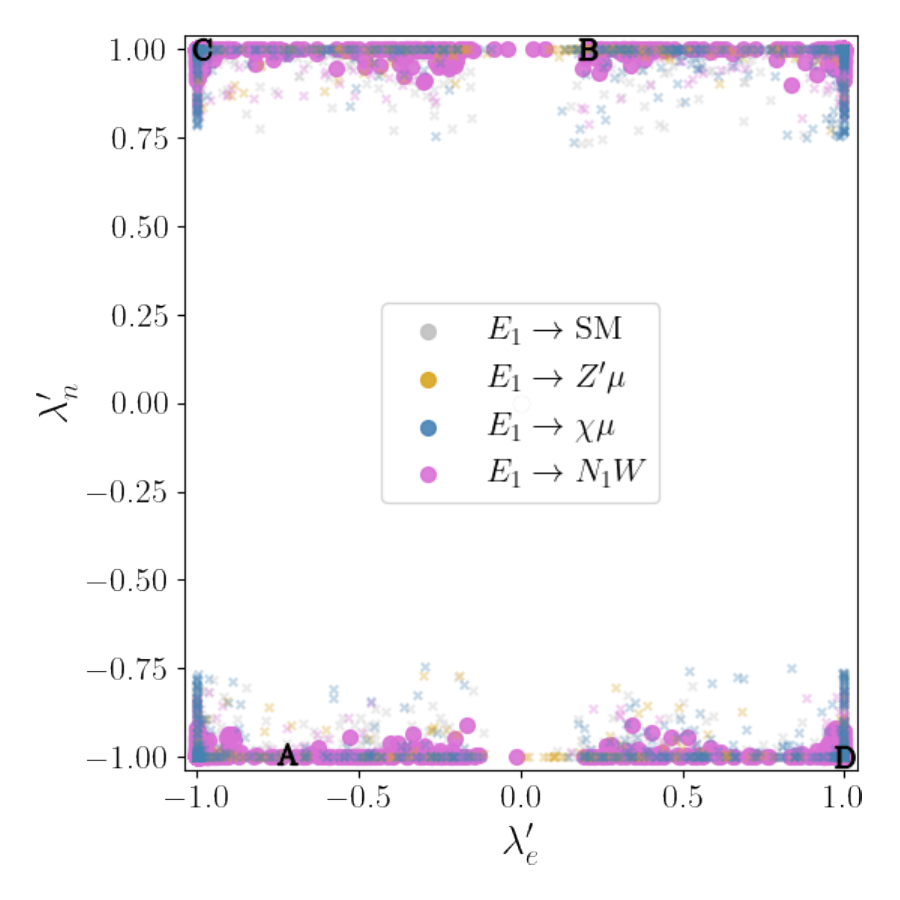}
\includegraphics[width=\fwidth]{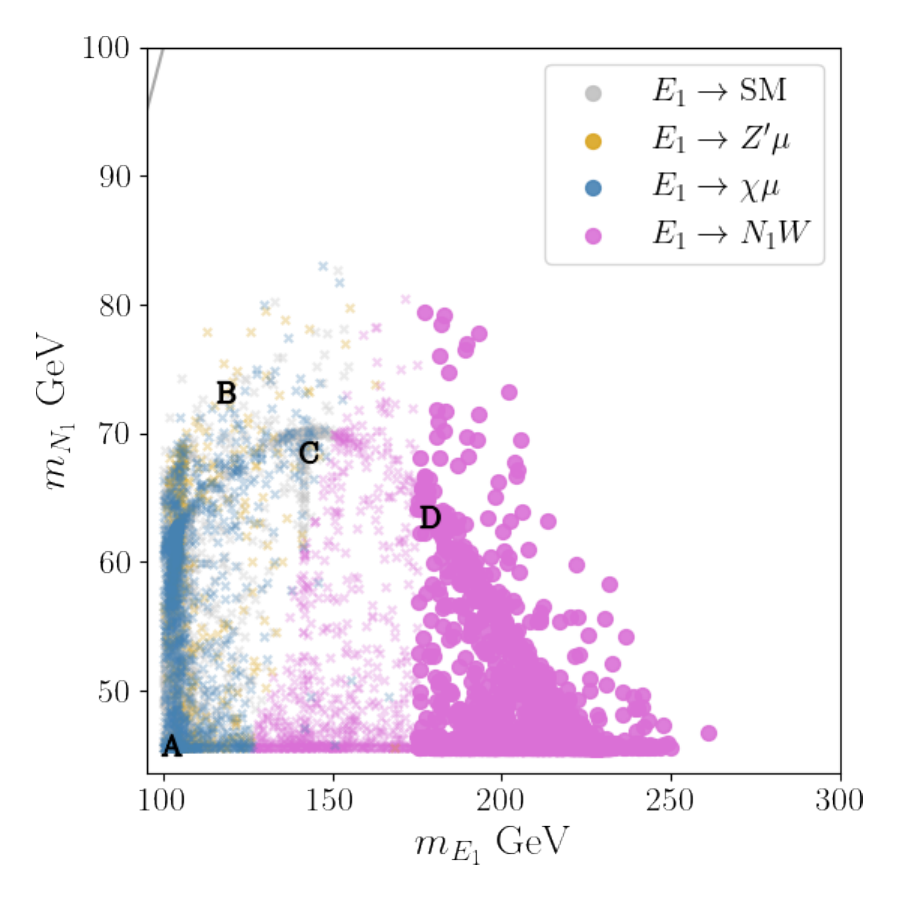}   \\
\caption{\label{fig-leln}
Scattering plots the Yukawa couplings $(\lpe,\lpn)$
and the VL lepton masses $(m_{E_1}, m_{N_1})$.
The cross and circle points are
$100 < m_{E_1} < 175~\GeV$ and $m_{E_1} > 175~\GeV$, respectively.
}
\end{figure}

\begin{figure}[tbh]
\centering
\includegraphics[width=\fwidth]{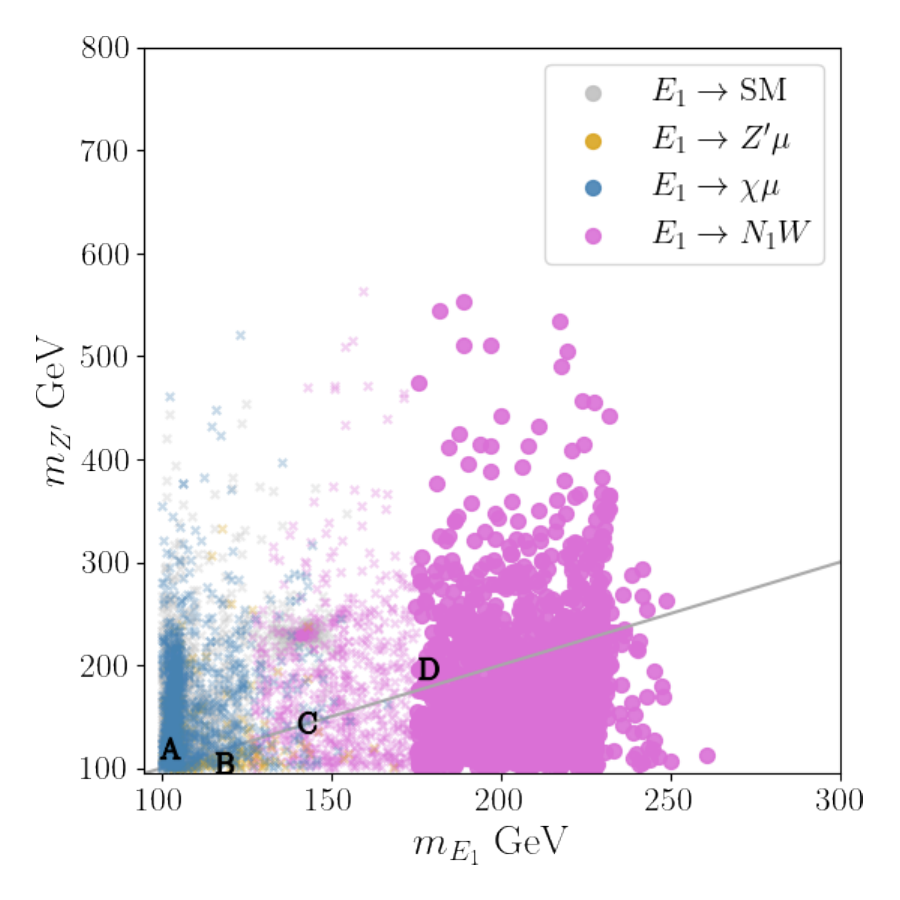}
\includegraphics[width=\fwidth]{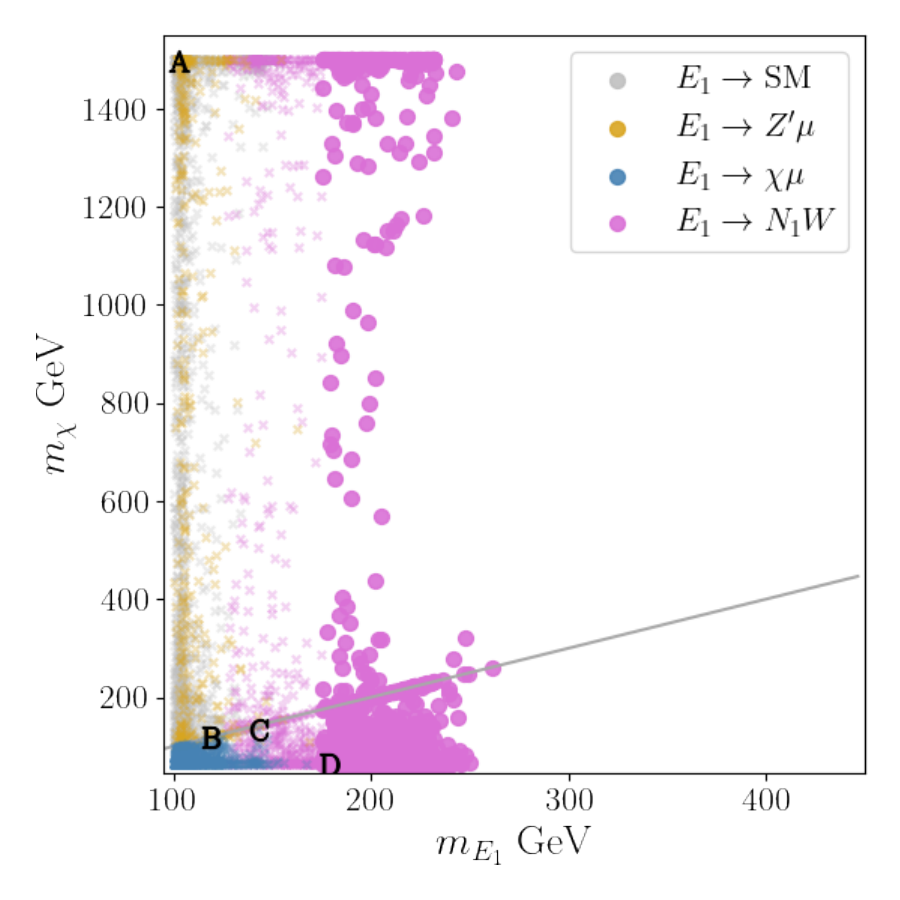}
\caption{\label{fig-mZpCh}
Scattering plots on the masses of $\Zp$ boson and the VL leptons.
The cross and circle points are
$100 < m_{E_1} <  175~\GeV$ and $m_{E_1} > 175~\GeV$, respectively.
The diagonal lines are respectively $m_{\Zp} = m_{E_1}$ and $m_{\chi} = m_{E_1}$
on the left and right panel.
}
\end{figure}

\begin{figure}[tbh]
\centering
\includegraphics[width=\fwidth]{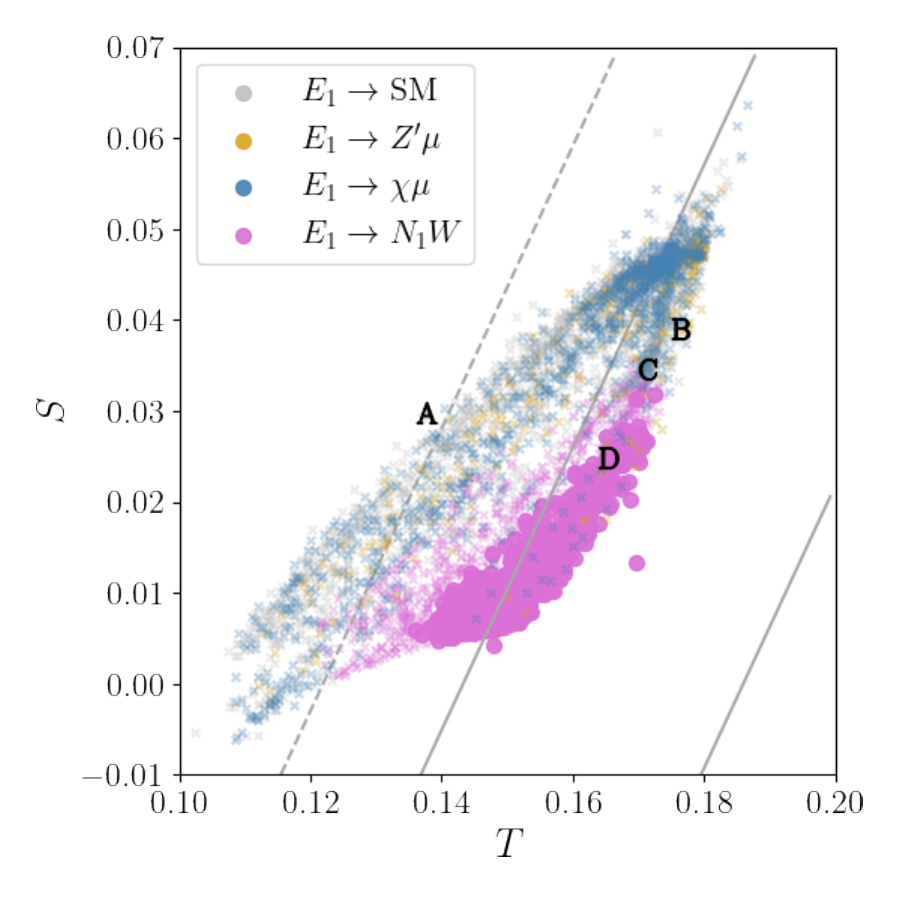}
\includegraphics[width=\fwidth]{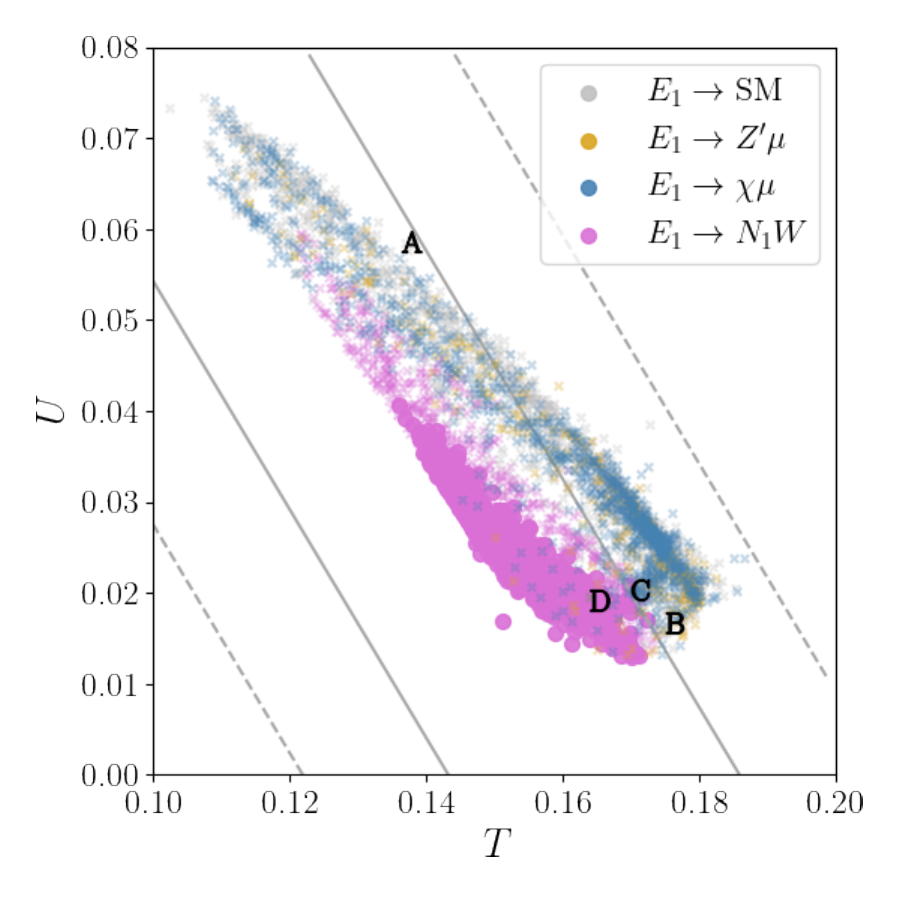}
\caption{\label{fig-STU}
Scattering plots on the oblique parameters.
The cross and circle points are
$100 < m_{E_1} < 175~\GeV$ and $m_{E_1} > 175~\GeV$, respectively.
The solid (dashed) lines are $1\sigma$ ($2\sigma$) range of the $W$ boson mass
measured by the CDF for $U=0$ and $S=0$ on the left and right panel, respectively.
}
\end{figure}

There are 12 parameters in our model,
\begin{align}
x = (\gp, v_\Phi, m_\chi, m_L, m_E, m_N, \la_L, \la_E, \la_e, \lpe, \la_n, \lpn),
\end{align}
to be scanned in our parameter search.
The SM Yukawa coupling constant $y_2$ is fixed to explain the muon mass
for a given parameter set.
We take $\la_N = y_n = 0$
since these are irrelevant for phenomenology after integrating out $\nu_R$.
To find points which can explain the anomalies consistently with the other observables,
we minimize the $\chi^2$ function
\begin{align}
 \chi^2(x) :=  \sum_I \frac{(y_I(x) - y^0_I)^2}{\sigma_I^2},
\end{align}
where $y_I$'s are observables listed in Table~\ref{tab-ex}. 
$y^0_I$ and $\sigma_I$ are the experimental central value and its $1\sigma$ error,
respectively.
For $R_\mathrm{CCFR}$, the error is chosen such that the $2\sigma$ deviation corresponds to the 95\% C.L limit.
Here, we assume that the $W$ boson mass is given by the CDF result.
In the SM, $\chi^2_\SM \simeq 94$ 
mainly originated from $\Delta a_\mu$ and $m_W$ 
as well as the $2.7\sigma$ deviation in $\Gam{W}{\tau\nu}/\Gam{W}{\mu\nu}$.

In our analysis, we restrict the parameter space to be
\begin{align}
& v_\Phi \in [100, 2000]~\GeV,
&\quad &
m_L, m_E \in [100, 1500]~\GeV, \\
& m_\chi \in [63, 1500]~\GeV
&\quad &
m_N \in [40, 1500]~\GeV,
\end{align}
and
\begin{align}
\gp \in [0, 0.35],
\quad
\la_L, \la_E, \la_e, \la_n, \lpe, \lpn \in [-1,1].
\end{align}
Here, the lower bounds of $m_\chi$ is chosen so that $h\to\chi\chi$
is kinematically forbidden.
The $U(1)^\prime$ gauge constant $\gp$ is required to be smaller than 0.35,
so that it is perturbative up to $10^{16}~\GeV$
in the full model with VL quarks~\cite{Kawamura:2019rth}.
We minimize the $\chi^2$ function from random initial points in the parameter range.
We further require $m_{E_1}, m_\Zp > 100~\GeV$ and $m_{N_1} > 45~\GeV$
for the fitted points,
so that the collider limits would be avoided and $Z\to N_1N_1$ is kinematically forbidden.

We calculated branching fractions of the $\Zp$ boson and the VL leptons.
For the $\Zp$ boson decay,
we calculated the two-body decays to the leptons.
We neglect the decays to quarks,
since the couplings to quarks will be tiny as discussed in Section~\ref{sec-bsll}.
For the decays of VL leptons,
we calculated the two-body decays to a SM, $\chi$, $\Zp$ boson and a lepton.
The three-body decays via off-shell $\Zp$ or $W$ boson
are calculated based on the formula shown in Appendix~\ref{sec-three}
if the corresponding on-shell two-body decay is kinematically forbidden.

Figures~\ref{fig-obs},~\ref{fig-leln},~\ref{fig-mZpCh} and~\ref{fig-STU} 
show the scatter plots of the result of the $\chi^2$ fitting.
On these plots, the cross and circle points are
$100 < m_{E_1} < 175~\GeV$ and $m_{E_1} > 175~\GeV$, respectively.
The colors of points indicate the dominant decay mode of the lightest charged VL
lepton $E_1$.
Note that the three-body decays are also classified by its off-shell boson,
e.g. $E_1 \to N_1 e \nu_e$ is a part of $E_1 \to N_1 W$ for $m_{E_1}-m_{N_1} < m_W$.

The scatter plots for the observables are shown in Fig.~\ref{fig-obs}.
On the left panel, the gray error bars are the $1\sigma$ errors along the axes.
We see that $\Delta a_\mu$ and $m_W$
are simultaneously explained on these points within $1\sigma$.
From the right panel, $R_{\gamma\gamma}$ can deviate from unity at most $4\%$,
and hence this is about $1\sigma$ below the experimental central value of $1.10$.
The neutrino trident process can be induced from the $\Zp$ boson exchange,
and there are points close to the upper bound of $1.10$,
depicted by the gray dashed line.
Altogether, all of the points explain the anomalies in $m_W$ and $\Delta a_\mu$
without conflicting with the current limits
on $R_{\gamma\gamma}$ and $R_{\mathrm{CCFR}}$.

\afterpage{\clearpage}

The left panel of Fig.~\ref{fig-leln}
shows the scatter plots on $(\lpe, \lpn)$.
There are points only where $|\lpn| \sim 1.0$.
This indicates that the mass shift of $m_W$
is predominantly induced by the VL neutrino which can be lighter than the charged ones.
The right panel shows the same plot on $(m_{E_1}, m_{N_1})$.
We see that the VL neutrino should be lighter than 80 GeV
to explain the shift of $m_W$,
and the upper bound becomes stronger for heavier $E_1$.
Thus, most of the points with $m_{E_1} > 175~\GeV$
have $E_1\to N_1 W$ as the dominant decay mode.

Figure~\ref{fig-mZpCh} shows the scatter plots on the $m_{E_1}$
and $m_\Zp$ ($m_\chi$) plane on the left (right) panel.
The gray line is $m_{E_1}= m_{\Zp}$ or $m_\chi$.
The upper bound on $m_\Zp$ is about $600~\GeV$
to explain the muon $g-2$, while $\chi$ can be much heavier.
$E_1\to N_1 W$ is the dominant decay mode even if $m_{\Zp} < m_{E_1}$
due to the larger gauge coupling constant $g > \gp$.

Figure~\ref{fig-STU} shows the scatter plots for the oblique parameters.
The solid (dashed) lines correspond to $1\sigma$~($2\sigma$)
errors of $m_W$ of the CDF measurement when $U=0$ and $S=0$ on the left and right panel,
respectively.
For $m_{E_1} > 175~\GeV$,
the points are found only around $(S,T,U) \sim (0.01, 0.15, 0.02)$,
so the $W$ boson mass shift is mainly explained by the $T$ parameter.
Different values of the oblique parameters are allowed
for $m_{E_1}< 175~\GeV$.
These patterns of the oblique parameters
will be useful to distinguish the parameter space of our model.

\begin{table}[t]
\centering
\caption{\label{tab-benchInp}
Selected input parameters and branching fractions of new particles. 
The observables not included in our $\chi^2$ analysis are shown in the last four rows. 
}
\begin{tabular}[t]{c|cccc}\hline
 benchmark & A & B & C & D \\ \hline \hline
$g^\prime$& 0.3479& 0.1212& 0.3500& 0.3500 \\ \hline
 $m_{Z^\prime}$~GeV & 119.2054& 105.5420& 145.2143& 196.6287 \\ 
 $m_\chi~\mathrm{GeV}$& 1498.6692& 118.1512& 135.2181& 63.1570 \\ 
 $m_{E_1}~\mathrm{GeV}$& 102.2595& 118.0921& 142.4606& 178.4233 \\ 
 $m_{E_2}~\mathrm{GeV}$& 322.9197& 156.0050& 1524.0154& 1517.7766 \\ 
 $m_{N_1}~\mathrm{GeV}$& 45.8666& 73.1853& 68.6071& 63.5020 \\ 
 $m_{N_2}~\mathrm{GeV}$& 216.9324& 247.9042& 245.9130& 258.8140 \\ \hline\hline
 $\mathrm{Br}(Z^\prime \to \mu\mu)$& 0.0057& 0.6488& 0.0017& 0.0033 \\ 
 $\mathrm{Br}(Z^\prime \to \nu\nu)$& 0.0000& 0.1059& 0.0002& 0.0006 \\ \hline
 $\mathrm{Br}(N_1\to \chi \nu)$& 0.0000& 0.0000& 0.0000& 0.9938 \\ 
 $\mathrm{Br}(N_1\to Z^\prime \nu)$& 0.0006& 0.9896& 0.0027& 0.0009 \\ 
 $\mathrm{Br}(N_1\to W \mu)$& 0.9994& 0.0104& 0.9973& 0.0054 \\  \hline 
 $\mathrm{Br}(E_1\to \chi \mu)$& 0.0000& 0.0000& 0.6454& 0.0163 \\ 
 $\mathrm{Br}(E_1\to Z^\prime \mu)$& 0.0010& 0.9999& 0.0000& 0.0000 \\ 
 $\mathrm{Br}(E_1\to W N_1)$& 0.0000& 0.0000& 0.0000& 0.9836 \\ 
 $\mathrm{Br}(E_1\to \mathrm{SM})$& 0.9990& 0.0001& 0.3546& 0.0000 \\ \hline\hline
 $S$& 0.0299& 0.0390& 0.0347& 0.0249 \\ 
 $T$& 0.1377& 0.1763& 0.1713& 0.1652 \\ 
 $U$&  0.0586& 0.0168& 0.0204& 0.0191 \\ 
 $R(V_{us})$&  1.0002& 1.0000& 1.0000& 1.0000 \\ 
 \hline  
 \end{tabular}
\end{table}

\begin{table}[t]
\centering
\caption{\label{tab-benchObs}
Values of observables at the benchmark points. 
The second column is the SM prediction. 
The degree of freedom is $15-12 = 3$. 
}
\begin{tabular}[t]{c|ccccc}\hline
 benchmark & SM & A & B & C & D \\ \hline \hline
$\chi^2$& 93.7831& 15.2403& 16.2259& 16.0397& 16.1686 \\ \hline
 $\Delta a_\mu$$\times 10^9$ & 0.0000& 2.4826& 2.5001& 2.5136& 2.5027 \\ 
 $m_W~\mathrm{GeV}$& 80.3610& 80.4338& 80.4335& 80.4338& 80.4334 \\ \hline
 $\mathrm{Br}(W\to\mu\nu)$& 0.1084& 0.1084& 0.1084& 0.1084& 0.1084 \\ 
 $\Gamma(Z\to \mu\mu)$& 0.0840& 0.0840& 0.0840& 0.0840& 0.0840 \\ 
 $\Gamma(Z\to \mathrm{inv})$& 0.4976& 0.4977& 0.4976& 0.4976& 0.4976 \\ 
 $A_\mu$& 0.1468& 0.1475& 0.1468& 0.1470& 0.1468 \\ 
 $A^\mu_{\mathrm{FB}}$& 0.0162& 0.0162& 0.0162& 0.0162& 0.0162 \\ 
 $R_{\mu\mu}$& 1.0000& 0.9277& 0.9999& 0.9693& 0.9955 \\ 
 $R_{\gamma\gamma}$& 1.0000& 1.0282& 1.0016& 1.0044& 1.0011 \\ \hline
 $\Gamma(Z\to\mu\mu)/\Gamma(Z\to ee)$& 1.0000& 0.9994& 1.0000& 0.9999& 1.0000 \\ 
 $\Gamma(Z\to\tau\tau)/\Gamma(Z\to \mu\mu)$& 1.0000& 1.0006& 1.0000& 1.0001& 1.0000 \\ 
 $\Gamma(W\to\mu\nu)/\Gamma(W\to e\nu)$& 1.0000& 1.0000& 1.0000& 1.0000& 1.0000 \\ 
 $\Gamma(W\to\tau\nu)/\Gamma(W\to \mu\nu)$& 1.0000& 1.0000& 1.0000& 1.0000& 1.0000 \\ 
 $\Gamma(\tau\to\mu\nu\nu)/\Gamma(\tau\to e\nu\nu)$& 0.9726& 0.9725& 0.9726& 0.9726& 0.9726 \\ \hline
 $R_\mathrm{CCFR}$& 1.0000& 1.0000& 1.0050& 1.0002& 1.0004 \\ 
\hline
 \end{tabular}
\end{table}

Tables~\ref{tab-benchInp} and~\ref{tab-benchObs} shows the benchmark points of our model.
The second row of Table~\ref{tab-benchObs} is the SM predictions 
whose $\chi^2$ value is about $94$, 
whereas the benchmark points have $\chi^2 \sim 15$. 
The improvement of $\chi^2 - \chi^2_\SM \sim 79$ 
is dominantly from $\Delta a_\mu$ and $m_W$.  
At the point (A), the Higgs boson decays can be changed by $7\%$ and $3\%$
in $h\to \mu\mu$ and $h\to\gamma\gamma$, respectively,  
due to the light VL lepton, $m_{E_1} \sim 100~\GeV$.  
The other observables, including 
$\Gamma(W\to \tau\nu)/\Gamma(W\to \mu\nu)$ which gives $2.7\sigma$, 
are not changed from the SM values. 
The values of the oblique parameters $S,T,U$ and $R(V_{us})$ 
are shown in the last four rows of Table~\ref{tab-benchInp} for reference.  
We find $R(V_{us})\gtrsim 1$ as expected from the analytical analysis.

The points (A), (B), (C) and (D) are chosen from the points
whose $E_1$ decay is dominated by
that to the SM particles, $\Zp\mu$, $\chi \mu$ and $N_1 W$, respectively.
Here, the SM particles include $Z\mu, h\mu$ and $W\nu$.
These points are plotted
on Figs.~\ref{fig-obs},~\ref{fig-leln},~\ref{fig-mZpCh} and~\ref{fig-STU}.
The branching fraction of the $\Zp$ boson
to the SM leptons are shown in this table.
The $\Zp$ boson will also decay to the VL leptons
if it is kinematically allowed~\footnote{
Signals same as the leptonic cascade decay $\Zp \to E_1 \mu$ at the LHC
is studied in Ref.\cite{Dermisek:2022xal}.
This search could constrain our parameter space
if the production cross section is sufficiently large.
}.
The lightest VL neutrino $N_1$ decays to a SM neutrino and a $\chi$ boson
if $m_\chi < m_{N_1}$ as in the point (D),
while it decays to a $W$ or $\Zp$ boson in the other cases.

\afterpage{\clearpage}

\subsection{LHC signals}

We found that $m_{E_1} \lesssim 250~\GeV$ is required to explain the shift of $m_W$.
Such a light VL lepton may be constrained by LHC searches, depending on its decay.

\renewcommand{\labelenumi}{(\Alph{enumi})}
\begin{enumerate}
 \item { $E_1 \to \SM$ \\
  The VL lepton would decay to a SM boson and a SM lepton.
  If $E_1$ is doublet-like,
  the latest limit is about $1000~\GeV$~\cite{CMS:2022nty}~\footnote{
We consider the limit from the search for the VL lepton pair production 
decaying to a tau lepton as a conservative limit for our VL lepton decaying to muon.  
}. 
  Even for the singlet-like case,  the limit is about 200 GeV~\cite{Dermisek:2014qca}
  using the run-1 data~\cite{ATLAS-CONF-2013-070}.
  Since $E_1$ is mostly doublet-like,
  the VL lepton lighter than $250~\GeV$ may be excluded by the current data.
  However, these decay modes may be suppressed
  because $[g^Z_{e_{L,R}}]_{E_1 \mu} = \order{m_\mu/m_{E_1}}$,
  and hence the other decay mode will coexist, or even dominate the decays.
  Actually, $E_1\to \SM$ is sizable only on the point (A). 
}
 \item{ $E_1 \to \Zp \mu$ \\
  In Ref.~\cite{Kawamura:2021ygg},
  it is shown that the limit for the doublet-like VL lepton is about 500 (1200) GeV
  for $\br{\Zp}{\mu\mu} = 2/3$ and $\br{E_1}{\Zp \mu} = 0.1~(1.0)$
  by the signal with four muons or more~\cite{ATLAS:2021yyr}.
  Although the $\Zp$ boson was assumed to be on-shell in Ref.~\cite{Kawamura:2021ygg},
  the limits on the VL lepton would be similar even for $m_{E_1} < m_{\Zp}$.
  At the point (B), this decay mode dominates the others
  and $\mathrm{Br}(E_1\to \Zp \mu\to \mu\mu\mu)\sim 0.65$,
  so this case may also be excluded.
}
\item{ $E_1 \to \chi \mu$ \\
  This is the dominant decay mode on the point (C).
  This case may not be excluded if $\chi$ decays to quarks.
  The $\chi$ coupling to two SM fermions is induced only by the fermion mass effects,
  so $\chi\to bb$ can be the dominant mode~\footnote{
  For $m_\chi > 2 m_t$, 
  $\chi \to tt$ may be the dominant decay mode,
  so the signal would be constrained by the 4 top quarks search~\cite{ATLAS:2018kxv}.
  Such heavy $\chi$ boson is, however, always heavier than $E_1$,
  and $E_1 \to \chi \mu$ is kinematically forbidden.
}.
  Another possibility is $\chi \to hh^*$
  via the quartic interaction $|H|^2 |\Phi|^2$ in the scalar potential.
  Since the decays to fermions are suppressed, this mode can dominate over the others.
  In this case, we expect $4h + \mu\mu$ from the $E_1$ pair production.
  Yet another possibility is
  loop-induced decays such as $\chi \to \gamma\gamma$ and $gg$,
  which may be particularly important for $m_\chi < m_h$.
  Thus there are so many decay modes that could be sizable
  and many parameters independent of our discussions which are involved. Thus this case may not be excluded.
}
\item{ $E_1 \to W N_1$ \\
  This decay mode may dominate the others when $m_{N_1} < m_{E_1}$
  which is favored to explain the $m_W$ shift.
  The VL neutrino $N_1$ then decays as $N_1 \to \Zp \nu$ or $N_1 \to \chi \nu$.
  At the point (D), $N_1\to \chi \nu$ dominates,
  so the signal is $E_1 E_1 \to WW + \chi\chi+ \nu\nu$.
  This signal contains many particles in the final state,
  but these will be relatively soft because of the smaller phase space.
  Thus, there may be a case in which the point (D) is not excluded by the current bounds.
}
\end{enumerate}

We note that the pair production of $N_1$
is also constrained by the $\ge 4\mu$ search~\cite{Kawamura:2021ygg}  
or the VL lepton search~\cite{CMS:2022nty}, as for $E_1$ pair production. 
Although $N_1$ is mostly singlet-like,
there is a mixing with the doublet-like state,
and thus $N_1$ can be pair produced from a $Z$ boson.
The light $N_1$ would be excluded
if $\br{N_1}{\Zp\nu}$ or $\br{N_1}{W\mu}$ is sizable.  
Hence, the point (C) would also be excluded by the direct $N_1$ search.
Therefore, the benchmark points other than the point (D) may be excluded
by the current data.
At the point (D), the decay of $\chi$ is important for the collider signals
of both $N_1$ and $E_1$.
As we have already discussed, there are various decay modes of the $\chi$ boson
because of the vanishing couplings to two SM fermions, see Eq.~\eqref{eq-Ychie},
and thus there may be cases which any search can not constrain.
This is an interesting subject, but is beyond the scope of this paper.

\section{Summary}
\label{sec-concl}

In this work, we studied the $W$ boson mass
in the SM extension with VL leptons and a $U(1)^\prime$ gauge symmetry, in
which only the VL leptons carry non-zero charges.
The full model with VL quarks was originally considered to explain the anomalies
in the muon $g-2$ and the $b\to s\ell\ell$ decay
simultaneously~\cite{Kawamura:2019rth,Kawamura:2019hxp}.
We explicitly studied the $W$ boson mass and $\Delta a_\mu$
induced by the VL leptons and the $\Zp$ boson,
and we found points which can explain both anomalies.
Since $b\to s\ell \ell$ can easily be explained by tiny couplings with the SM quarks,
as discussed in Sec.~\ref{sec-bsll},
our model provides a simultaneous explanation for the three anomalies.

The VL leptons should be light to explain the shift of $m_W$.
We found that the lightest charged (neutral) VL lepton $E_1$ ($N_1$)
is lighter than 250 (80) GeV
if absolute values of the Yukawa couplings are less than unity.
In our model, the VL leptons typically decay through the $\Zp$ or $\chi$ boson.
If the VL lepton decays through the $\Zp$ boson, which may decay to di-muons,
there will be strong constraints from the searches
for signals with four muons or more~\cite{Kawamura:2021ygg}.
Hence, the decay to the $\chi$ boson should dominate the decay of the VL lepton,
which may be achieved by the mass hierarchy $m_\chi < m_{E_1,N_1} < m_{\Zp}$.
The $\chi$ boson decays to the SM particles in various ways,
and thus there may be some cases for which none of the LHC searches exclude our model
with the light VL leptons.
Studying the decays of the $\chi$ boson in the full model with VL quarks is our future work.

\section*{Acknowledgment}
The work of J.K.
is supported in part by
the Institute for Basic Science (IBS-R018-D1),
and the Grant-in-Aid for Scientific Research from the
Ministry of Education, Science, Sports and Culture (MEXT), Japan No.\ 18K13534.
The work of S.R. is supported in part by the Department of Energy (DOE) under Award No.\ DE-SC0011726.

\appendix

\section{Diagonalization of mass matrices} 
\label{sec-nonzero}

We first discuss the diagonalization of the charged lepton mass matrix Eq.~\eqref{eq-Me}. 
The diagonalization matrix is decomposed as
\begin{align}
 U_L = U_{L}^0 U_{L}^1 U_{L}^2,
\quad
 U_R = U_{R}^0 U_{R}^1 U_{R}^2.
\end{align}
The unitary matrices are given by
\begin{align}
 U_{L}^0 =
\begin{pmatrix}
 c_L & s_L & 0 \\  -s_L & c_L  & 0 \\ 0 & 0 & 1
\end{pmatrix},
\quad
 U_{R}^0 =
\begin{pmatrix}
 c_E & s_E & 0 \\  -s_E & c_E  & 0 \\ 0 & 0 & 1
\end{pmatrix},
\end{align}
\begin{align}
U_{L}^1 =&\
\begin{pmatrix}
 1- y_{42}^2 \dfrac{v_H^2}{2M_E^2} &
y_{22} y_{24} \dfrac{v_H^2}{M_L^2} - y_{42} \lpe  \dfrac{v_H^2}{M_E M_L}  &
 y_{42}\dfrac{v_H}{M_E} \\
- y_{22} y_{24} \dfrac{v_H^2}{M_L^2} + y_{42} \lpe \dfrac{v_H^2}{M_E M_L}  &
1 & 0 \\
- y_{42}\dfrac{v_H}{M_E} & 0 &  1- y_{42}^2 \dfrac{v_H^2}{2M_E^2}
\end{pmatrix}
+ \order{m_\mu^3},  \\
U_{R}^1 =&\
\begin{pmatrix}
 1- y_{24}^2 \dfrac{v_H^2}{2M_L^2} &
y_{22} y_{42} \dfrac{v_H^2}{M_E^2} - y_{24} \lpe \dfrac{v_H^2}{M_E M_L}  &
 y_{24}\dfrac{v_H}{M_L} \\
- y_{22} y_{42} \dfrac{v_H^2}{M_E^2} + y_{24} \lpe \dfrac{v_H^2}{M_E M_L}  &
1 & 0 \\
- y_{24}\dfrac{v_H}{M_L} & 0 &  1- y_{24}^2 \dfrac{v_H^2}{2 M_L^2}
\end{pmatrix}
+ \order{m_\mu^3},
\end{align}
and
\begin{align}
 U_{L}^2 =
\begin{pmatrix}
 1 & 0 & 0 \\ 0 & c_{e_L} & -s_{e_L} \\ 0 &  s_{e_L} & c_{e_L}
\end{pmatrix},
\quad
 U_{R}^2 =
\begin{pmatrix}
 1 & 0 & 0 \\ 0 & s_{e_R} & c_{e_R} \\ 0 & c_{e_R} & -s_{e_R}
\end{pmatrix},
\end{align}
where
\begin{align}
\label{eq-cseLR}
 \begin{pmatrix}
 s_{e_R} & c_{e_R} \\
 c_{e_R} & -s_{e_R}
\end{pmatrix}
\begin{pmatrix}
 y_{44} v_H & M_E \\ M_L & \lpe v_H
\end{pmatrix}
 \begin{pmatrix}
 c_{e_L} & -s_{e_L} \\
 s_{e_L} & c_{e_L}
\end{pmatrix}
 = \mathrm{diag}\left(m_{E_1}, m_{E_2} \right). 
\end{align}
Here,
$y_{22} = c_L c_E y_2+ s_L s_E \lae$,
$y_{24} = c_E s_L y_2 - s_E c_L \lae$,
$y_{42} = c_L s_E y_2 - c_E s_L \lae$
and
$y_{44} = s_L s_E y_2 + c_L c_E \lae$.
The unitary matrices $U_{L,R}^{1}$ block diagonalize the SM
and VL leptons up to $\order{y_2^2 v_H^2}$ when we assume $\la_e \sim \order{y_2}$.
These are simply identity matrices if we neglect $\order{m_\mu/v_\Phi}$.
The unitary matrices $U_{{L,R}}^0$ keeps the $SU(2)_L$ gauge couplings unchanged,
while $U_{{L,R}}^1$ do change the couplings as
\begin{align}
\label{eq-UL1}
\left(U_{L}^1\right)^\dag
Q_L
 U_{L}^1
=&\
\begin{pmatrix}
 1-y_{42}^2 \dfrac{v_H^2}{M_E^2} & 0 & y_{42} \dfrac{v_H}{M_E} \\
0 & 1 & 0 \\
y_{42} \dfrac{v_H}{M_E} & 0 & y_{42}^2 \dfrac{v_H^2}{M_E^2}
\end{pmatrix},
\quad
\left(U_{R}^1\right)^\dag Q_R U_{R}^1
=
\begin{pmatrix}
y_{24}^2 \dfrac{v_H^2}{M_L^2} & 0 & - y_{24} \dfrac{v_H}{M_L} \\
0 & 0 & 0 \\
- y_{24} \dfrac{v_H}{M_L} & 0 & 1 - y_{24}^2 \dfrac{v_H^2}{M_L^2}
\end{pmatrix}.
\end{align}
For $\lae v_H \lesssim m_\mu$,
the correction to the muon coupling deviates from the SM value within $\order{10^{-6}}$
for $\order{100}~\GeV$ VL lepton masses,
while it can be $\order{1}$ for  $\la_e \sim \order{1}$
with the fine-tuning for $y_{22} v_H \sim m_\mu$.
Note that $U_{e_{L,R}}^2$ does not change the SM coupling
since these rotate only the VL block.

The neutrino mass matrix, Eq.~\eqref{eq-Mn}, can be diagonalized as follows.
First, we rotate $\Mcal_n$ as
\begin{align}
\widetilde{\Mcal}_n =  \Mcal_n V_L^0 =
\begin{pmatrix}
 m_D & \cdot & \cdot \\
 0 & 0 & \tm_N \\
 0 & \tM_L & m_{n} \\
\end{pmatrix},
\quad
V_L^0 =
\begin{pmatrix}
 1 & 0 & 0 \\ 0 & c_N & s_N \\ 0 & -s_N & c_N
\end{pmatrix}
\begin{pmatrix}
 \tc_L & \ts_L & 0 \\ -\ts_L & \tc_L & 0 \\ 0 & 0 & 1
\end{pmatrix}
,
\end{align}
where $m_D= \tc_L y_n v_H + \ts_L s_N \la_N v_\Phi$,
$\tm_N = \sqrt{m_N^2 + \la_n^2 v_H^2}$,
$\tm_L = c_N m_L - s_N \lpn v_H$, $m_n = s_N m_L + c_N \lpn v_H$
and $\tM_L = \sqrt{\tm_L^2 + \la_L^2 v_\Phi^2}$.
The mixing angles are defined as
\begin{align}
 c_N := \frac{m_N}{\tm_N},
\quad
 s_N := \frac{\la_n v_H}{\tm_N},
\quad
 \tc_L := \frac{\tm_L}{\tM_L}
\quad
 \ts_L :=  \frac{\la_L v_\Phi}{\tM_L}.
\end{align}
The dots in the first low are linear combinations of $y_n v_H$ and $\la_N v_\Phi$,
which are irrelevant for our discussion
after integrating out the right-handed neutrino $\nu_R$.
The active neutrino mass is given by $m_D^2/M_R$.
The unitary matrices to diagonalize the neutrino mass is given by
\begin{align}
 U_{n_L} = V_L^0
\begin{pmatrix}
1 & 0 & 0 \\
0 & c_{n_L} & -s_{n_L} \\
0 & s_{n_L} & c_{n_L}
\end{pmatrix},
\quad
 U_{n_R} =
\begin{pmatrix}
1 & 0 & 0 \\
0 & s_{n_R} &  c_{n_R} \\
0 & c_{n_R} & -s_{n_R}
\end{pmatrix},
\end{align}
up to $v_\Phi/M_R \ll 1$, where
\begin{align}
 \begin{pmatrix}
 s_{n_R} & c_{n_R} \\
 c_{n_R} & -s_{n_R}
\end{pmatrix}
\begin{pmatrix}
 0 & \tm_N \\ \tM_L & m_n
\end{pmatrix}
 \begin{pmatrix}
 c_{n_L} & -s_{n_L} \\
s_{n_L} & c_{n_L}
\end{pmatrix}
 = \mathrm{diag}\left(m_{N_1}, m_{N_2} \right)
\end{align}
With the unitary matrix $U_{n_L}$,
the SM muon neutrino coupling to the $Z$ boson is rescaled as
$c_L^2 + c_N^2 s_L^2$.
Therefore, $s_N \sim \la_n v_H/m_N \ll 1$ is required
for the SM-like $Z$ boson coupling to muon neutrinos.

\section{Decay widths}
\label{sec-three}

We calculate the three-body decay, $F\to f_3 V^* \to f_3 \ol{f}_2 f_1$,
of fermions via a off-shell vector boson $V$ whose couplings are given by,
\begin{align}
 \Lcal = V_\mu \ol{f}_1 \gamma^\mu \left(g_L^{12} P_L + g_R^{12} P_R \right)  f_2
        + V_\mu \ol{f}_3 \gamma^\mu \left(g_L^{3F} P_L + g_R^{3F} P_R \right)  F.
\end{align}
The partial width is given by
\begin{align}
&\ \Gamma\left(F\to f_3 \ol{f}_2 f_1 \right)
=   \frac{m_{F}}{768\pi^3}
\left(\abs{g_L^{12}} ^2+\abs{g^{12}_R}^2\right)  \\ \notag
&\ \times \int^{(1-\sqrt{y})^2}_0 dt
\frac{\beta(t)}{(t-z)^2}
 \left[
  \left( \abs{g_L^{3F}}^2  + \abs{g_R^{3F}}^2\right)
 \left\{(1-y)^2 + (1+y)t-2t^2 \right\}
- 12 \mathrm{Re}\left(g_L^{3F} g_{R}^{3F*} \right)
  \sqrt{y}\; t
  \right],
\end{align}
where $\beta(t) = \sqrt{t^2-2(1+y)t+(1-y)^2}$,
$y=m_3^2/m_F^2$ and $z = m_V^2/m_F^2$.
Here, $m_{3}$, $m_F$ and $m_V$ are respectively the masses of $f_3$, $F$ and $V$,
and we neglect the masses of the fermions $f_1$ and $f_2$.

%%%%%%%%%%%%%%%%%%%%%%
%\clearpage
{\small
\bibliographystyle{JHEP}
\bibliography{reference_vectorlike,ref_fermionportal}
}

\end{document}